\documentclass[5p, twocolumn, nonatbib]{elsarticle}
\makeatletter %
\let\c@author\relax
\def\ps@pprintTitle{%
    \let\@oddhead\@empty
    \let\@evenhead\@empty
    \def\@oddfoot{\footnotesize\itshape
         }%
    \let\@evenfoot\@oddfoot
    }
\makeatother

\usepackage{silence}
\ErrorFilter{latex}{Command \bibfont already defined}

\usepackage{float} 

\usepackage{graphicx} %
\usepackage{subfigure} %

\usepackage{amsmath}
\usepackage{mathtools}
\usepackage{amssymb} %
\DeclareMathOperator*{\argmax}{arg\,max}
\DeclareMathOperator*{\argmin}{arg\,min}

\usepackage[backend=biber, natbib=true, useprefix=true, style=ieee, maxcitenames=1, mincitenames=1]{biblatex}

\AddToHook{env/frontmatter/after}{\setcounter{author}{0}}
\journal{}

\usepackage{pgfplots}
\usepackage{pgfplotstable} %
\pgfplotsset{compat=1.18}
\usepackage{tikz}
\usepackage{dblfloatfix}

\usepackage{booktabs}

\usepackage{adjustbox}
\usepackage{orcidlink} 
\begin{document}

\begin{frontmatter}

\title{Uncertainty Quantification in Machine Learning for Biosignal Applications - \\{A Review}}

\author{Ivo Pascal de Jong\orcidlink{0000-0002-1497-8013}\fnref{fn1}\corref{cor1} }
\ead{ivo.de.jong@rug.nl}

\author{Andreea Ioana Sburlea\orcidlink{0000-0001-6766-3464}\fnref{fn1}}

\author{Matias Valdenegro-Toro\orcidlink{0000-0001-5793-9498}\fnref{fn1}}

\affiliation[fn1]{organization={Department of Artificial Intelligence, Bernoulli Institute, University of Groningen},
city={Groningen},
country={The Netherlands}}

\cortext[cor1]{Corresponding author}

\begin{abstract}
\textbf{Purpose:} Uncertainty Quantification (UQ) has gained traction in an attempt to improve the interpretability and robustness of machine learning predictions. Specifically (medical) biosignals such as electroencephalography (EEG), electrocardiography (ECG),  electrooculography (EOG), and electromyography (EMG) could benefit from good UQ, since these suffer from a poor signal-to-noise ratio, and good human interpretability is pivotal for medical applications. To determine how uncertainty estimation can be used for biosignal tasks, we investigate current methods, use cases, applications, evaluations, and uncertainty measures.

\noindent\textbf{Methods:} In this paper, we systematically review the state of the art of applying Uncertainty Quantification to Machine Learning tasks in the biosignal domain. All works from Web of Science, Scopus, IEEE XPlore and PsycINFO that discuss uncertainty in Machine Learning on one of the aforementioned biosignals is included. 

\noindent\textbf{Results:} We present various methods, shortcomings, uncertainty measures and theoretical frameworks that currently exist in this application domain based on the 53 reviewed papers and related literature. We address misconceptions in the field, provide recommendations for future work, and discuss gaps in the literature in relation to diagnostic implementations as well as control for prostheses or brain-computer interfaces. 

\noindent\textbf{Conclusion:} Overall it can be concluded that promising UQ methods are available, but that research is needed on how people and systems may interact with an uncertainty-model in a (clinical) environment. 
\end{abstract}

\begin{keyword}
Uncertainty Quantification \sep Bayesian Neural Networks \sep Biosignals \sep EEG \sep ECG  \sep EMG %
\end{keyword}

\end{frontmatter}

\section{Introduction}
Standard Machine Learning (ML) systems such as Random Forests, SVMs, and Neural Networks typically produce single-point estimates for their classification task. Such single-point models neglect alternative predictions that are consistent with the training data, and therefore give an inadequate estimate of the uncertainty of a prediction. As a result, they may give overconfident but completely inaccurate predictions, which induces skepticism and hinders the implementation of Machine Learning methods in clinical settings \cite{he2019practical}. Uncertainty Quantification (UQ) attempts to address this problem by adapting Machine Learning systems to also predict a measure of confidence for a given prediction. Over the past years this has been gaining traction in Computer Vision \cite{abdar2021review}, but it is still only lightly explored in Machine Learning tasks that focus on Biosignals.

Applications using biosignals can gain particular benefits from uncertainty quantification. Their signals are sensitive to artifacts that could corrupt the prediction of a Machine Learning system in unexpected ways. Uncertainty Quantification methods may help here by recognizing that the data is corrupted and indicate increased uncertainty.

Another argument for the importance of Uncertainty Quantification is that the human interpretation of the signal requires substantial time investment. Automating this work with a Machine Learning model requires UQ to indicate when the model does not know and minimise misclassifications. To give an order of scale to the human effort: sleep scoring a patient's EEG recording of an overnight stay will typically take a neurologist about two hours \cite{malhotra2013performance}. A Machine Learning system that can automatically classify the majority of the overnight stay with high confidence while identifying the parts that it is uncertain on may reduce the amount of work for the neurologist.

\begin{figure*}[!htb]
\centering
\subfigure[Uncertainty for Decision Support. The uncertainty, prediction and data are available for the clinician. This requires interpretable uncertainties.]{
  \begin{minipage}[b]{0.45\linewidth}
  \centering
  \includegraphics[width=\linewidth]{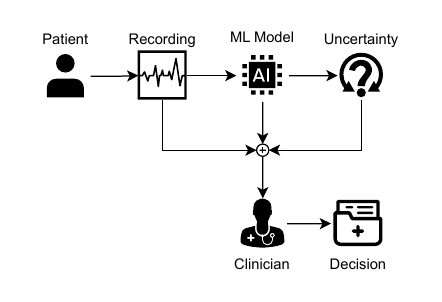}
  \end{minipage}
}\hspace{0.05\linewidth}
\subfigure[Uncertainty for Rejection. The model will make a decision if it is highly certain and likely to be correct. Otherwise, the data is given to the clinician. This requires uncertainty that can separate highly accurate predictions, and reduces diagnostic workload on the clinician. ]{
  \begin{minipage}[b]{0.45\linewidth}
  \centering
  \includegraphics[width=\linewidth]{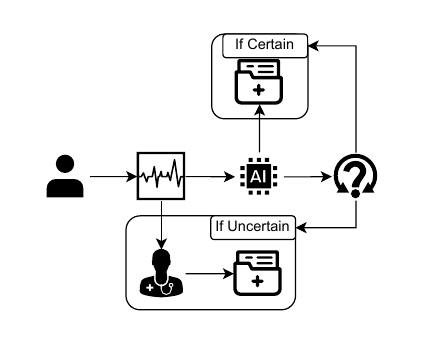}
  \end{minipage}
}
\subfigure[Uncertainty for improved ML. Uncertainty may be used in methods to improve the classification performance. This puts no direct restrictions on the uncertainty.]{
  \begin{minipage}[b]{0.45\linewidth}
  \centering
  \includegraphics[width=\linewidth]{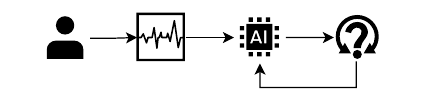}
  \end{minipage}
}\hspace{0.05\linewidth}
\subfigure[Uncertainty for extra recordings. Additional or alternative tests may be run when the model is uncertain. The predicted uncertainty should align with a clinician's uncertainty. ]{
  \begin{minipage}[b]{0.45\linewidth}
  \centering
  \includegraphics[width=\linewidth]{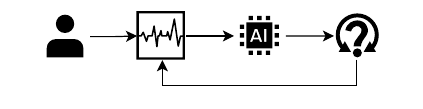}
  \end{minipage}
}
\caption{Different positions of Uncertainty Quantification in medical biosignal interpretation. Different ways of using uncertainty put different constraints on the predicted uncertainties. These designs are not mutually exclusive and uncertainty estimation may deliver multiple benefits. }
    \label{fig:uq_biosignal_diagram}
\end{figure*}

Figure \ref{fig:uq_biosignal_diagram} shows various roles uncertainty estimation can play in a biosignal Machine Learning system. The primary use cases are to improve transparency of predictions for a decision support system, or to make independent classifications only when it is likely to be correct. Additionally, uncertainty estimates may be used in various ways to improve the predictions of a Machine Learning model, and it may even be used to determine when additional medical tests are needed. The interactions with a clinical system put specific expectations on uncertainty estimation for biosignal applications.%

With the value that this direction of research can bring
this review attempts to identify how Uncertainty Quantification methods should be used in biosignal applications. Answering this question directly is impossible, but by investigating and critically assessing the way research is currently being conducted we provide some adjustments to the current directions and suggest new avenues to be explored in the future. Moreover, we provide an overview of currently common methods as an entryway for researchers new to the topic of UQ in Biosignal processing, together with a simplified end-to-end guide for implementing, applying and evaluating uncertainty.

In the rest of this section we explain how the literature review was performed to offer some usability, and we end the section with a thorough explanation of what uncertainty is. In Section \ref{sec:UQ-Methods} we discuss different methods for quantifying uncertainty. For each method we specifically discuss the relation to biosignal tasks, and we discuss niche methods that were used in biosignal tasks but that are otherwise not considered in general Uncertainty Quantification review papers. 

In Section \ref{sec:UQ_measures} we address misconceptions and confusion we observed about how a numerical measure of uncertainty should be extracted from a predicted distribution generated by some of the uncertainty quantification methods. We discuss the different uncertainty measures encountered, and give clear recommendations. While this topic is discussed a bit in the most cited review on Uncertainty Quantification \cite{abdar2021review}, we provide a more explicit overview and comparison, including insights from recent research on uncertainty measures. 

Then, Section \ref{sec:uq-usecases} we describe different ways uncertainty has been used in the biosignal domain. We discuss how the choice of use case is important as it affects what properties it should have and how it should be evaluated. This is not always apparent. By giving these guidelines we intend to make it clearer for authors and reviewers what uncertainty is useful for and how that should be evaluated.

We conclude our review paper with two sections that aim to progress the research on uncertainty in biosignals. In Section \ref{sec:how_to_build_uq} we provide a guideline on how uncertainty quantification may be added to a biosignal classifier, and in Section \ref{sec:open-challenges} we discuss open research challenges for applying uncertainty in biosignals. Those challenges focus specifically on the interaction of an uncertainty estimating model with the environment in which it is deployed, specific properties of biosignal data, and more broadly how uncertain Machine Learning behaves in a clinical setting.

\subsection{Literature Review Method}
To ensure reproducibility this section documents the systematic review based on the PRISMA 2020 guidelines \cite{page2021prisma}. We searched for papers matching the following query in their title, abstract or keywords:
\begin{align*}
 ((\text{Uncertainty  Quantification}& \land \text{Machine Learning}) \\
 &       \lor \text{Bayesian Neural Networks}) \\ 
 & \land \text{Biosignals}
\end{align*}

Conceptually, this means all papers that mention Uncertainty Quantification, Machine Learning, and Biosignals, or only Bayesian Neural Networks and Biosignals. \footnote{We found that a line of research \cite{chai_channels_2017, rifai_chai_classification_2016, rifai_chai_comparing_2015} uses the term "Bayesian Neural Networks" to describe classical Neural Networks trained with Bayesian Regularization \cite{burden2009bayesian}. These are not included. }

\begin{figure}[t]
    \centering
    \includegraphics[width=\columnwidth]{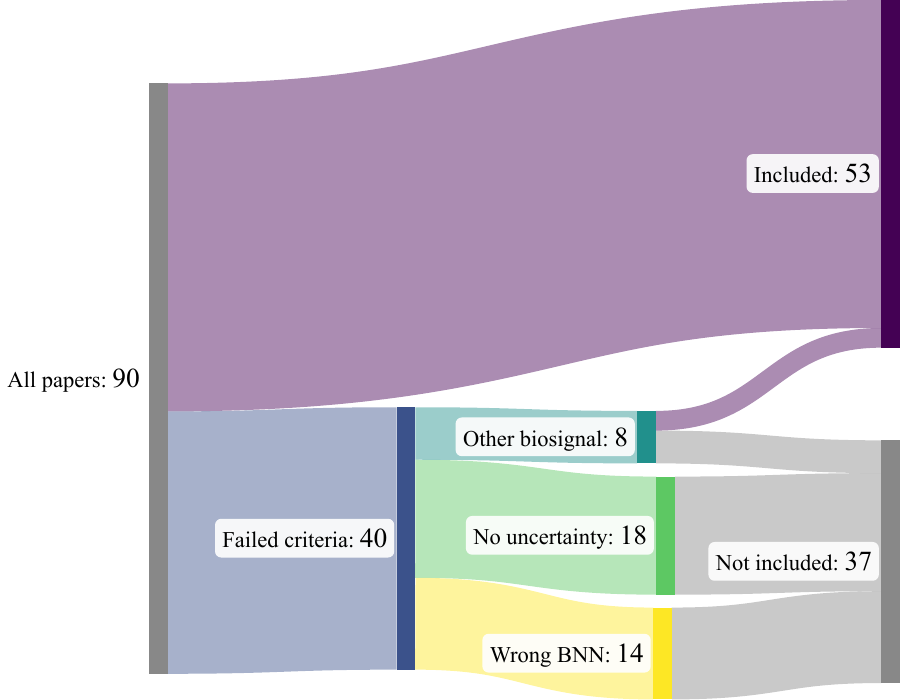}
    \caption{The flow of papers that were covered in the systematic literature search, divided by exclusion criteria.}
    \label{fig:flowchart_papers}
\end{figure}

To ensure good coverage of the review, various synonyms and abbreviations were used for each term. Specifically for the ``Machine Learning" term several Neural Networks methods were used (``Neural Networks", ``Convolutional Neural Networks", ``Deep Belief Networks", ``Hopfield Networks" and ``Multi-Layer Perceptrons"), and various Machine Learning models: ``Support Vector Machine", ``Random Forests", ``Linear Discriminant Analysis" and ``Fuzzy Logic". For the biosignals we searched on the following terms: ``EEG", ``ECG", ``EOG", ``EMG", ``BCI" and ``fNIRS". The choice of these terms was selected for the consistent modality, as each of them covers data from a set of time series from different sensor locations. For the concept of uncertainty we used the terms ``uncertainty quantification", ``uncertainty calibration", ``aleatoric uncertainty", ``epistemic uncertainty". ``uncertainty estimation", ``uncertainty model", ``uncertainty aware". For all terms the abbreviations, singular, and plural variations are included. 

The search was applied to the databases: Web of Science, Scopus, IEEE~Xplore and PsycINFO. The last search was conducted in January of 2025, including all work before 2024.

\definecolor{purple-black}{HTML}{190129}
\definecolor{purple0}{HTML}{440154}
\definecolor{blue1}{HTML}{3B528B}
\definecolor{blue2}{HTML}{21908C}
\definecolor{green3}{HTML}{5DC863}
\definecolor{yellow4}{HTML}{FDE725}

\begin{figure}[t]
    \resizebox{\columnwidth}{!}{%

        \pgfplotstableread[row sep=crcr]{
Label	EEG	    ECG	    sEMG	MRI      EOG     fNIRS\\
2005	1.00	0.00	0.00	0.00     0.00    0.00\\
2007	1.00	0.00	0.00	0.00     0.00    0.00\\
2018	2.00	0.00	0.00	0.00     0.00    0.00\\
2019	0.00	1.00	0.00	0.00     0.00    0.00\\
2020	0.00	3.00	0.00	0.00     1.00    0.00\\
2021	2.00	6.00	0.00	1.00     0.00    1.00\\
2022	5.00	1.00	1.00	1.00     0.00    0.00\\
2023	7.00	11.00	3.00	0.00     0.00    0.00\\
        }\testdata
    
        \begin{tikzpicture}
    
        \begin{axis}[
            ybar stacked,
            ylabel=Count,
            ymin=0,
            ymax=25,
            xtick=data,
            legend style={cells={anchor=west}, legend pos=north west},
            reverse legend=false, %
            xticklabels from table={\testdata}{Label},
            xticklabel style={text width=2cm,align=center},
            xlabel=Year
        ]
        \addplot+[color=yellow4]   table [y=EEG, meta=Label, x expr=\coordindex] {\testdata};
        \addlegendentry{EEG}
        \addplot+[color=green3]  table [y=ECG, meta=Label, x expr=\coordindex] {\testdata};
        \addlegendentry{ECG}
        \addplot+[color=blue2]  table [y=sEMG, meta=Label, x expr=\coordindex] {\testdata};
        \addlegendentry{EMG}
        \addplot+[color=blue1]  table [y=MRI, meta=Label, x expr=\coordindex] {\testdata};
        \addlegendentry{MRI}
        \addplot+[color=purple0] table [y=EOG, meta=Label, x expr=\coordindex] {\testdata};
        \addlegendentry{EOG}
        \addplot+[color=purple-black] table [y=fNIRS, meta=Label, x expr=\coordindex] {\testdata};
        \addlegendentry{fNIRS}
        \addplot [
            ybar, %
            nodes near coords,
            nodes near coords style={%
                anchor=south,%
            },
        ] table [ y expr=0.00001, x expr=\coordindex] {\testdata};
    
        \end{axis}
        \end{tikzpicture}
    
    }
    \caption{Histogram of the number of papers per year using Uncertainty Quantification for each Biosignal. Overall this shows an increase in the number of Biosignal papers using Uncertainty Quantification. This shows an increase in the popularity of Uncertainty Quantification methods.}
    \label{fig:yearplot}

\end{figure}
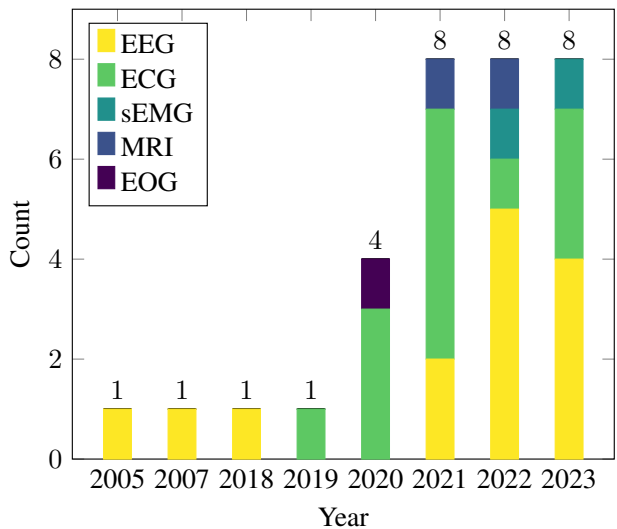

Only papers that investigate Machine Learning with Uncertainty Quantification on one of the selected biosignals are considered. One person manually collected all papers from the search that based on the abstract and title could meet this criteria, yielding a total of 90 papers. Reviewing the full papers showed that of these 90 papers, 50 met the criteria. 14 papers used the ``Bayesian Neural Networks" term to refer to a regularization method instead of uncertainty estimation method, 18 did not look at uncertainty of an ML model, and 8 papers did not concern a relevant biosignal. Another three papers looked at different biosignals, but were kept due to their interesting application of uncertainty quantification. The number of included and not included papers as well as their exclusion criteria are visualised in Figure \ref{fig:flowchart_papers}. 

From each of the remaining 53 papers, one person determined: the considered biosignal, the task, the Machine Learning method, the type of Machine Learning task, the Uncertainty Quantification method(s), the uncertainty quantification use case, how the uncertainty is evaluated, the uncertainty measures used and the dataset used. Additional notes were taken about promising future directions, critical interpretation of the work, overarching themes and limitations, and further details about the data processing. This forms the basis of the synthesis, with a strong emphasis on the overarching themes, gaps in current research, and guidelines for sound implementation of uncertainty estimation methods. 

This review approach does not aim to be fully deterministic and purely data driven, but instead aims to generate a full exposition of the current body of research, and generate insights and critical perspectives that help advance the fields and streamline entry for other researchers. Access to the data and notes is available for reproducibility upon request.

Figure \ref{fig:yearplot} shows an overview of the results from this search. It shows that from 2018 to 2023 there has been an increase in the use of Uncertainty Quantification. This shows a growing interest in applying Uncertainty Quantification to the biosignal domain.

\subsection{Fundamentals of Predictive Uncertainty}\label{sec:fundamental_uncertainty}

Before going into the specific Machine Learning models that can quantify uncertainty for a given prediction, it is important to first understand what uncertainty really entails. \citet{hullermeier2021aleatoric} explains how predictive uncertainty can arise from two conceptual sources: aleatoric uncertainty and epistemic uncertainty. In the biosignal literature various definitions are used, some of which are incomplete. We give a thorough and exact definition of both and add clarifications.

Aleatoric uncertainty\footnote{Aleatoric is derived from the Latin word "alea", meaning "dice" or "chance".} is the uncertainty that comes from stochasticity in the true function $f: X \rightarrow y$ from which dataset $D$ is sampled. This means that aleatoric uncertainty arises when the optimal function given infinite samples still does not perfectly predict $y$. 

From this definition follows that aleatoric uncertainty cannot be reduced by having a better model, and that humans also cannot give better predictions. Even with arbitrarily many training samples, the aleatoric uncertainty will not decrease. Aleatoric uncertainty is commonly simplified to either label noise (such as imperfect annotations) or sensor noise in the inputs. Artifacts that destroy the underlying signal such as disconnected leads or signal clipping cause aleatoric uncertainty at the inputs. 

Epistemic uncertainty\footnote{Epistemic is derived from the Greek word "episteme", meaning "knowledge".} (also known as model uncertainty) is the uncertainty that comes from not knowing the true function $f: X\rightarrow y$. The learned model $f^\theta$ may not match the true model due to model misspecification, limited approximation quality, or limited training samples. 

Under epistemic uncertainty a better model or a better human expert would be able to make a more accurate prediction. Epistemic uncertainty may arise when a model is applied to data that is different from the data it was trained on, which is referred to as out-of-distribution \cite{yang2021generalized}. It may also arise when there are not enough training samples to learn a good approximation of $f$. The exact boundary between epistemic uncertainty due to out-of-distribution samples or due to limited in-distribution samples is not clear. Therefore, it can be easier to treat both of these conditions as epistemic uncertainty. Unlike aleatoric uncertainty, epistemic uncertainty does decrease with an increase in training samples. 

Artifacts that obscure the signal such as baseline drift or line noise make learning the true function harder, but not impossible. Therefore, these are sources of epistemic uncertainty. Epistemic uncertainty will arise in biosignals when there is insufficient (diverse) training samples. If a classifier is to be applied on different people, different hardware, or in different contexts this introduces generalisation error, which is caused by epistemic uncertainty.

\begin{figure}[!tb]
\centering
\subfigure[Predictions with Aleatoric ]{
  \begin{minipage}[t]{0.45\linewidth}
  \centering
  \includegraphics[width=\linewidth]{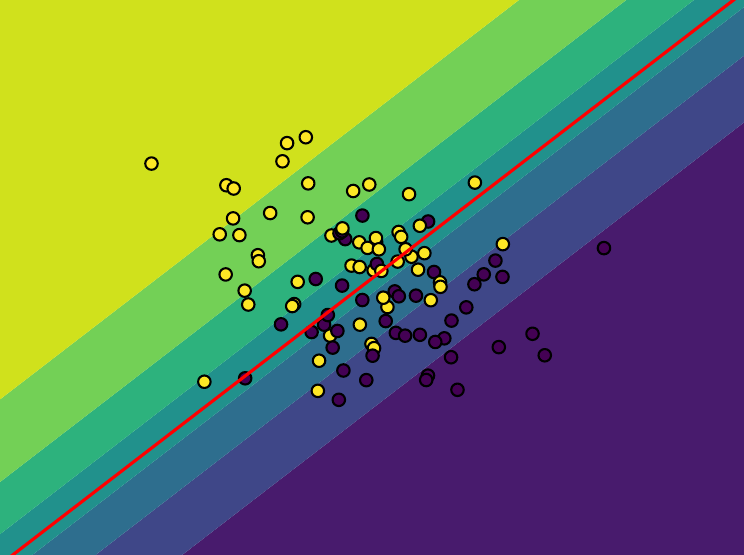}\label{fig:ale}
  \end{minipage}
}
\subfigure[Prediction with Epistemic]{
  \begin{minipage}[t]{0.45\linewidth}
  \centering
  \includegraphics[width=\linewidth]{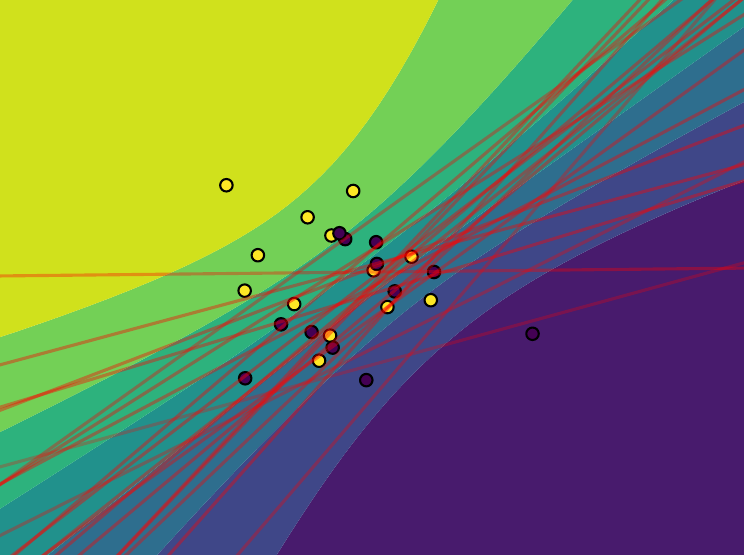}\label{fig:epi}
  \end{minipage}
}
\caption{Predictions with aleatoric or epistemic uncertainty in the 2D feature space of a binary classification task. The dots represent training samples, and the background colour the uncertain predictions. In Figure \ref{fig:ale} aleatoric uncertainty is shown as two classes for which the distribution of the data overlaps, resulting in uncertain predictions in the background. Epistemic uncertainty is shown in Figure \ref{fig:epi} where due to limited data it is not clear which decision boundary (red) is the true decision boundary.}
\label{fig:uncertain_classification}
\end{figure}

The distinction between aleatoric and epistemic uncertainty is made clear in Figure \ref{fig:uncertain_classification}, which shows how aleatoric and epistemic uncertainty arise in classification. In this case we see that in the area of feature space where both classes occur, aleatoric uncertainty arises. Epistemic uncertainty arises as the model cannot perfectly learn the distribution of the classes in feature space. 

\Citet{van_gorp_certainty_2022} emphasises the need for this distinction in sleep stage classification, although this need also applies to other areas. They explain how aleatoric uncertainty should be addressed differently than epistemic uncertainty. If there is high aleatoric uncertainty for a given ECG sample, theoretically there would be no use in having a clinician review the same ECG for a second opinion as they would not be able to give a better prediction. Instead you should consider getting another recording, or collecting additional information. \citet{larsen2023new} for example proposes to run SPECT-MPI tests only when an ECG classifier is uncertain to create a multi-stage classifier. In practice, because aleatoric uncertainty is estimated by an imperfect model it is very possible that a clinician would be able to make a better prediction.

For epistemic uncertainty more (relevant) training data, better models, or having the samples interpreted by a clinician can improve the quality of a diagnosis. 

\subsubsection{Limitations of Aleatoric and Epistemic Uncertainty }

In Section \ref{sec:UQ-Methods} we will discuss how aleatoric and epistemic uncertainty can present differently in some ML methods, and discuss methods that claim to be able to separate them. However, we first want to highlight the limitations of estimating aleatoric and epistemic uncertainty. 

The primary limitation is that we currently cannot adequately quantify aleatoric and epistemic uncertainty separately in classification. In later sections we will introduce methods for quantifying aleatoric and epistemic uncertainty, but theoretical arguments \cite{wimmer2023quantifying}, observations \cite{mucsanyi2024benchmarking} and experimental demonstrations \cite{de2024disentangled} have shown that there are interactions between aleatoric uncertainty and epistemic uncertainty in classification. \citet{mucsanyi2024benchmarking} has shown that predictions of aleatoric and epistemic uncertainty are highly rank correlated, \citet{wimmer2023quantifying} has shown that under high aleatoric uncertainty current methods will not be able to predict epistemic uncertainty, and \citet{de2024disentangled} shows that this problem extends to multiple datasets, UQ methods, and uncertainty measures. While estimates of aleatoric and epistemic uncertainty may be useful, with the current method we cannot trust that a prediction of a certain kind of uncertainty is truly attributable to that specific uncertainty for classification. This makes the idea of different actions for different kinds of uncertainty as proposed in \cite{van_gorp_certainty_2022} infeasible with the current methods. 

Additionally, specifically in biosignal applications we should explicitly consider the role of preprocessing. Using fewer features or more aggressive filtering trades epistemic uncertainty for aleatoric uncertainty from the model's perspective. We therefore need to be aware and explicit in what we define as our learning task for which disentangled uncertainty is estimated. 

Since the aleatoric-epistemic perspective is only a perspective on uncertainty there are other ways to look at uncertainty. Some of these alternatives fit into the aleatoric-epistemic framework, but others do not. For example, in Section \ref{sec:edl} we discuss Prior Networks, where the epistemic uncertainty is split into \textit{epistemic uncertainty} and \textit{distributional uncertainty}. Meanwhile \citet{bishop2006pattern} makes a distinction between \textit{discriminative} and \textit{generative} models, where the former learns a decision boundary between the classes, and the latter learns the class likelihood in feature space. Under these generative models samples with low likelihood for either class may be considered uncertain. However, this does not intuitively fit into either aleatoric or epistemic uncertainty.

\subsubsection{Uncertainty in Terms of Evidence}

One alternative perspective on uncertainty is discussed in the literature. \citet{lin_reliability_2022}, distinguishes between uncertainty from \textit{vacuity} and from \textit{dissonance}. This comes from the domain of Subjective Logic \cite{josang2016subjective}. Here, vacuity is the absence of evidence for a prediction. Dissonance arises from conflicting evidence. \citet{lin_reliability_2022} describes these in a context of evidence-based Machine Learning. Similar to the aleatoric and epistemic uncertainty one can use this distinction to make decisions on how to improve the quality of a model.

This perspective is much less explored in the biosignal literature but warrants further research as it may be more suitable for interpretation by clinicians than the aleatoric-epistemic perspective.

\subsubsection{Conformal Prediction}
Another perspective on Machine Learning with uncertainty is Conformal Prediction. In Conformal Prediction, instead of outputting the most likely prediction with a measure of how uncertain that prediction is, we instead accept a \textit{set} of outputs such that we can be certain (up to some degree) that the correct answer is in the set. 

Formally we say that Conformal Predictions gives a set of answers $S$, such that $P(\hat{y} \in S) \geq 1-\alpha$, where $\alpha$ might be $5\%$, so that there's a $\geq95\%$ chance the correct estimate is one of the estimates. For some samples the model might give one (correct) answer, for others it might give several answers. In the common practical implementation (called Inductive Conformal Prediction), which answers should be included in the set $S$ is determined by setting a threshold against any measure of uncertainty. This threshold is found by learning the quantiles of a separate calibration dataset, which should be identically distributed to the test set. Conformal Prediction guarantees \textit{coverage}, meaning that $P(\hat{y} \in S) \geq 1-\alpha$ holds. Ideally, Conformal Prediction gives minimal sets to maintain coverage. A poor uncertainty scoring function will still have coverage, but the output set might include many possible answers all the time. 

Conformal Prediction may be easier to interpret by clinicians in a decision support system \cite{folgado2025conformal}, because it allows clinicians to equally consider multiple options. This is especially relevant in multi-class tasks such as arrhythmia classification. Though not found with the systematic search, there is some literature already studying Conformal Prediction on biosignal data.  \citet{eliades2018detecting} studies Conformal Prediction in EEG seizure detection and in EMG exoskeleton control \citet{eliades2019applying}, while giving a clear demonstration of how to evaluate a Conformal Prediction model.  \citet{wang2021speech} studies Conformal Prediction for decoding speech from sEMG using a BLSTM. They propose the use of Conformal Prediction in risk-sensitive situations, as it gives strong guarantees on safety. They use the Conformal Predictions to create labels for unlabelled samples whenever the model is certain, such that the model can be re-trained on a larger dataset.

\section{Methods for Uncertainty Quantification}\label{sec:UQ-Methods}

As most of the development of Uncertainty Quantification methods happens in the field of Computer Vision \cite{abdar2021review}, it is no surprise that the Machine Learning models for which Uncertainty Quantification is defined are models that perform well in Computer Vision. As a result we find most works build on Neural Networks. Specifically, this review found many Convolutional and Recurrent Neural Networks. An overview of the type of different Neural Network types is given in Figure \ref{fig:Neural Network models}.

With the vast majority of models being Neural Networks, the Uncertainty Quantification methods are also mostly intended for Neural Networks. An overview of the most common methods covered is given in Table \ref{tab:uq_methods_overview}. This gives a quick reference of the most important properties, but how the methods work and how they specifically relate to biosignals is discussed below. At the of this section Table \ref{tab:uq_methods_overview} gives a complete list of each method and the reviewed papers that use them.

This section mostly discusses Neural Networks methods for Uncertainty Quantification, as this is most extensively studied. First, the concept of Bayesian Neural Networks is explained, including the range of different implementations. Bayesian Neural Networks are the current standard for Uncertainty Quantification, and they lend themselves well to interpretation through the lens of aleatoric and epistemic uncertainty.  Next, we will discuss some other common Uncertainty Quantification methods such as Variational Autoencoders \cite{kingma2019introduction}, Evidential Deep Learning \cite{sensoy2018evidential} and Gaussian Process Regression \cite{costabal_machine_2019}. We also discuss post-hoc uncertainty calibration methods \cite{guo2017calibration}, and end this section with a list of the less established and novel methods for uncertainty quantification that have been used for biosignals. Altogether, this section gives a complete overview of the Uncertainty Quantification methods that are used in the biosignal application domain. 

\definecolor{purple0}{HTML}{440154}
\definecolor{blue1}{HTML}{3B528B}
\definecolor{blue2}{HTML}{21908C}
\definecolor{green3}{HTML}{5DC863}
\definecolor{yellow4}{HTML}{FDE725}

\begin{figure}[t]
    \resizebox{\columnwidth}{!}{%
        
        \begin{tikzpicture}
        \begin{axis}[
            ybar stacked,   
            ylabel={Count}, 
            legend style={cells={anchor=west}, 
                            legend pos=north east},
            symbolic x coords={CNN, CRNN,RNN,Transformer,MLP,VAE},
            xticklabel style={rotate=45, anchor=north east},
            bar width=0.2cm,
            xtick pos=left,
            ytick pos=left,
            ymin= 0 ,
        ]
        \addplot+[color=purple0]
        	coordinates {
            (CNN, 22) 
            (CRNN, 5) 
            (RNN, 3)
            (Transformer, 0)
            (MLP, 4)
            (VAE, 1)

            };
        \addlegendentry{Without attention}

        \addplot+[color=yellow4]
            coordinates {
                (CNN, 3)
                (CRNN, 1)
                (RNN, 0)
                (Transformer, 3)
                (MLP, 0)
                (VAE, 0)
            };
            
        \addlegendentry{With attention}

        \end{axis}
        \end{tikzpicture}
            
    }
    \caption{Popularity of various Neural Network architectures in this review. Models with at least one convolutional or recurrent layer are respectively labeled CNN or RNN. Models with both are labeled as CRNN. Yellow indicates models with attention layers.}
    \label{fig:Neural Network models}

\end{figure}
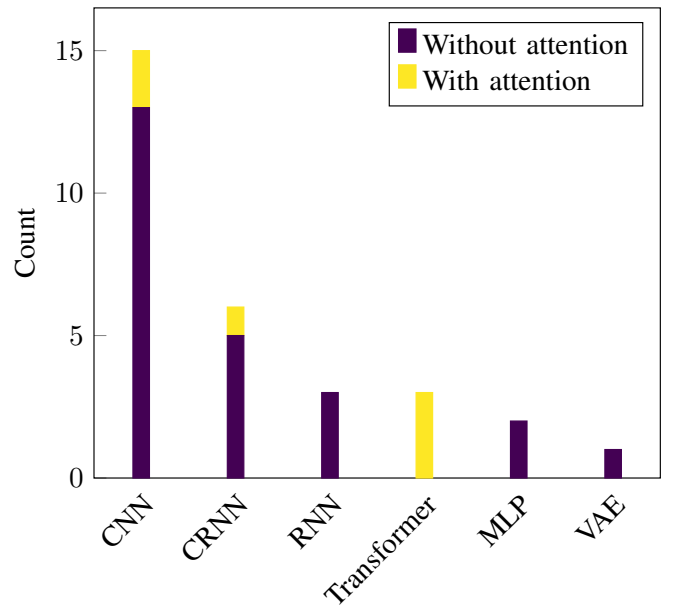

\subsection{Notation for Softmax Uncertainty}\label{sec:notation}

Standard Neural Networks give point-estimate predictions for a given sample. In regression, this prediction is a scalar with no indication of uncertainty or expected error. However, in classification with standard Neural Networks a Softmax activation function is often used such that the prediction is given as 
\begin{equation}\label{eq:softmax}
    p(y\,{=}\,c\,|\, x, \theta) = \frac{\exp(f_c^{\theta}(x))}{\sum_c' \exp(f_{c'}^{\theta}(x))}.
\end{equation}
Where $f_c^{\theta}$ predicts the logits for a given input $x$, as parameterized by $\theta$. To ease notation we introduce the predicted probability of a class $c$ as
\begin{equation}
    p_c \coloneqq p(y\,{=}\,c\,|\, x, \,D), 
\end{equation} which in the case of a standard Neural Network with parameters $\theta$ learned on dataset $D = \{\mathbf{X}, \mathbf{y}\}$ is approximated with $p_c = p(y\,{=}\,c\,|\, x,\theta)$.

Before going into how uncertainty is modelled in Bayesian Neural Networks, it is important to be aware that predicting class probabilities, rather than directly predicting a class label already quantifies uncertainty. However, it only quantifies aleatoric uncertainty and neglects epistemic uncertainty, making it overconfident under epistemic uncertainty.

We found a common misconception in the literature that normal Neural Networks cannot estimate predictive uncertainty. Using the predicted class probabilities they can, but possibly not very well. Therefore, applications of Bayesian Neural Networks for estimating uncertainty should consider an equivalent normal Neural Network as a baseline to justify the added complexity and computational cost.

\subsection{Heteroscedasticity}\label{sec:heterosedastic_BNN}

The above formulation for Softmax gives simple estimates for aleatoric uncertainty in the standard approach for classification with Neural Networks. Such class probabilities are standard in classification tasks, but not in regression. Standard regression models will only predict the best value, and not give any indication of uncertainty. In those models, uncertainty can still be derived from performance metrics like the Mean Squared Error. This assumes \textit{homoscedastic uncertainty}, where the risk of error is uniform throughout the feature space.

\begin{figure}[!tb]
\centering
\subfigure[Homoscedastic Regression]{
  \begin{minipage}[t]{0.45\linewidth}
  \centering
  \includegraphics[width=\linewidth]{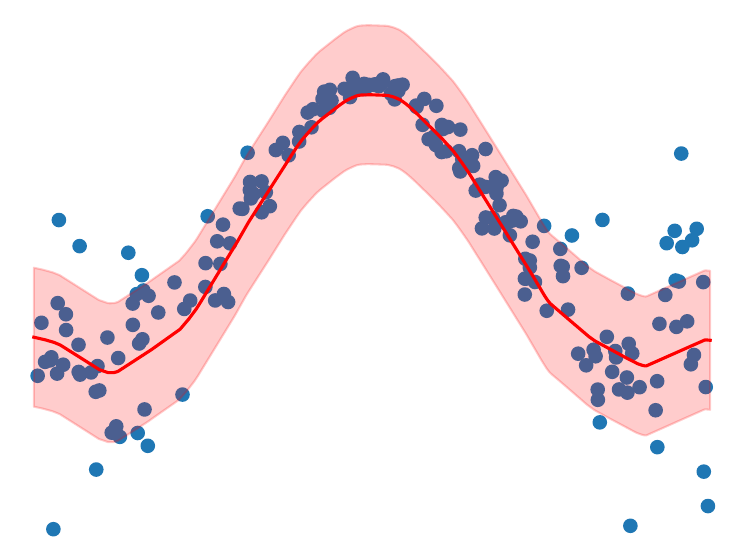}\label{fig:homo_regression}
  \end{minipage}
}
\subfigure[Heteroscedastic Regression]{
  \begin{minipage}[t]{0.45\linewidth}
  \centering
  \includegraphics[width=\linewidth]{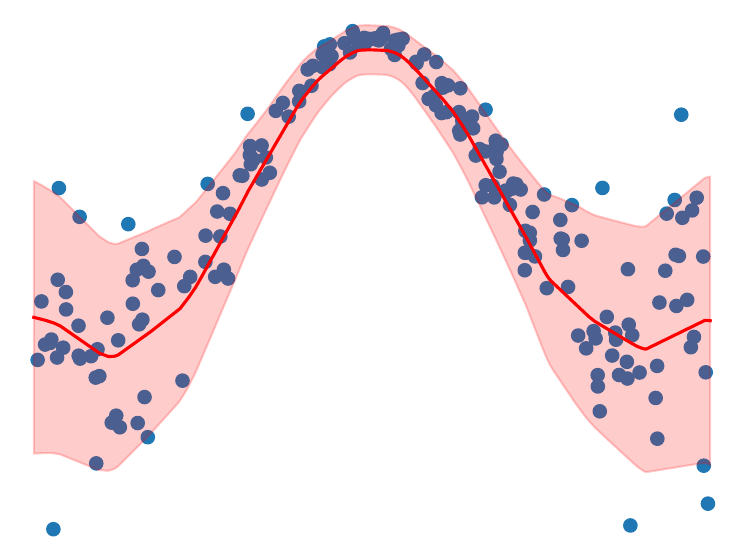}\label{fig:hetero_regression}
  \end{minipage}
}
\caption{Modelling homoscedastic and heteroscedastic uncertainty in regression. The dots indicate the training samples, the red line the regression prediction, and the red shaded areas the 95\% confidence interval based on either Mean Absolute Error (homoscedastic) or predicted absolute error (heteroscedastic). In both cases the data has heteroscedastic noise, but the predictions change on whether the models assume homoscedasticity or heteroscedasticity.}
\label{fig:homoscedastic_heteroscedastic_regression}
\end{figure}

To be able to distinguish between more and less difficult samples we need to consider \textit{heteroscedastic uncertainty}. In regression this can be done by having a second prediction that estimates the variance. In these models, the Neural Network not only attempts to learn a predicted regression value, but it has a separate output for the predicted error. This results in a prediction, paired with a measure of aleatoric uncertainty. Taking $\mu_\theta(x)$ as the predicted mean and $\sigma^2_\theta(x)$ as the predicted variance for a sample, the predicted value $\hat{y}$ is given as
\begin{equation}
    \hat{y} = \mathcal{N}(\mu_\theta(x), \sigma^2_\theta(x)).
\end{equation}
Such a model is then trained with a loss function that optimizes both the predicted mean and the variance. The Gaussian Negative Log-Likelihood loss 
\begin{equation}
    \mathcal{L}_{NLL}(y_{true}, x) = \frac{\log \sigma^2_{\theta}(x)}{2} + \frac{(\mu_\theta(x) - y_{true})^2}{2 \sigma^2_\theta(x)}
\end{equation}
is the simplest, but alternatives losses have been proposed \cite{seitzer2022pitfalls}. \citet{vranken2021uncertainty} and \citet{jin2023uncertainty} use this structure of mean and variance estimation in classification, to learn aleatoric uncertainty over logits.

Other approaches to estimating aleatoric uncertainty in regression are by estimating a lower and upper bound \cite{khosravi2010lower}, or estimating intervals without assuming any distribution \cite{betancourt2021interval}. Alternatively, heteroscedastic uncertainty may be estimated with Quantile Regression methods \cite{koenker2001quantile, jantre2021quantile}, where regression lines are learned for higher and lower quantiles.  Figure \ref{fig:homoscedastic_heteroscedastic_regression} shows the difference between homoscedastic and heteroscedastic uncertainty estimation in regression. It shows that if some parts have more or less noise in the output, then homoscedastic uncertainty estimation averages these out, whereas heteroscedastic uncertainty estimation maintains the distinction

\begin{figure}[t]
    \centering
    \includegraphics[width=\linewidth]{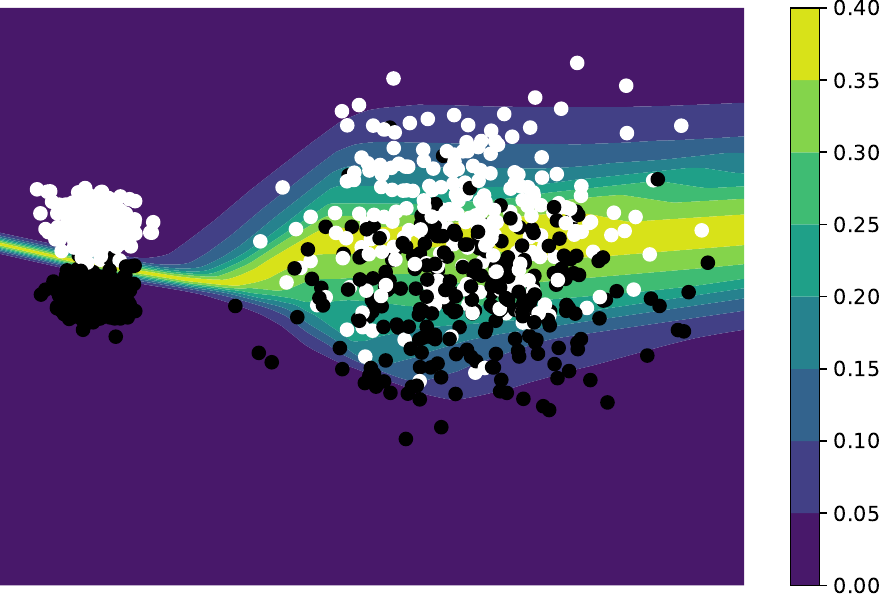}
    \caption{Heteroscedastic uncertainty in classification. The white and black dots represent samples from different classes. Background indicates heteroscedastic aleatoric uncertainty (entropy). %
    }
    \label{fig:heteroscedastic_classification}
\end{figure}

In Figure \ref{fig:heteroscedastic_classification} we show a classification problem with heteroscedastic uncertainty. The white dots represent one class, and the black dots another. At the cluster on the left these are well separated, with low uncertainty, but at the cluster on the right these overlap with high uncertainty. A simple multi-layer perceptron with Softmax shows increased uncertainty (bright background) where the class distributions overlap.

\subsection{Bayesian Neural Networks}\label{sec:bnn}

Given a starting point of aleatoric uncertainty with softmax, we move towards quantifying epistemic uncertainty with Bayesian Neural Networks. The foundational difference is the way both methods look at learning the parameters. In the standard Neural Network the parameters $\theta$ are learned from the space of all possible sets of parameters $\Theta$ to minimize a loss function $\mathcal{L}(\theta, D)$. The loss function primarily measures the error between the predictions and the annotated ground truth. Under Bayesian Neural Networks, instead of considering a single optimized set of parameters $\theta$, we consider a distribution of all possible sets of parameters in $\Theta$. Since some parameters are more likely under dataset $D$ than others, we also consider the likelihood of each set of parameters. This results in the integral
\begin{equation}\label{eq:aleatoric_epistemic}
p_c = \int \underbrace{p(y\,{=}\,c\,|\, x, \theta)}_{\text{Aleatoric}} \underbrace{p(\theta | D)}_{\text{Epistemic}} d\theta.
\end{equation}
From this the epistemic uncertainty as the probability distribution of the parameter vector $p(\theta | D)$ also becomes apparent. %

Some approximations of Bayesian Neural Networks such as MC-Dropout \cite{gal2016dropout} and Deep Ensembles \cite{lakshminarayanan2017simple} are based on this equation. They sample multiple parameter vectors $\theta$ which are all trained to maximise $p(\theta |D)$ through e.g. the negative log-likelihood. From each parameter vector predictions are made. The disagreement between these predictions now captures epistemic uncertainty.

To complete the picture of the Bayesian Neural Network, we take the dataset $D$ as 
 Random Variables $\{X, Y\}$ and deconstruct the posterior $p(\theta|D)$ with Bayes theorem as
\begin{equation}\label{eq:bayes_rule}
p(\theta| X, Y) = \frac{p(Y|X, \theta) p(\theta)}{p(Y|X)}.
\end{equation}
The evidence term $p(Y|X)$ is intractable\footnote{This would result in an integral for each parameter of the Neural Network such that ${p(Y|X) = \int_{\theta_1}\int_{\theta_2}\ldots\int_{\theta_D}p(Y|X, \theta) d\theta_1 d\theta_2 \ldots d\theta_D}$ where $D$ represents the number of dimensions of the parameter vector $\theta$.}. Fortunately, it is constant for a given dataset, so we can optimize $\theta$ only on the likelihood and the prior. The likelihood is determined by the model fit to the data and may be computed through a loss function. The prior $p(\theta)$ can be selected to match assumptions about the modelling task. %

The rest of this section explains different ways in which Bayesian Neural Networks are approximated to be computationally feasible. For each method we will provide a conceptual understanding, and show the specific limitations in how they might affect biosignal applications. 

\subsubsection{MC-Dropout}

Dropout~\cite{srivastava2014dropout} has been a prominent regularization method in Deep Learning applications. During training with dropout, some nodes have a probability $p$ to be dropped (i.e. activation set to 0). This adds noise to the training procedure and has been thoroughly shown to be an effective regularizer. 

Normally, the dropout is removed during inference to prevent dropping important information. MC-Dropout (Monte Carlo Dropout)~\cite{gal2016dropout} keeps this dropout during inference, and uses multiple predictions to make sure all important information is sampled. %

Dropout can be considered as a special probability distribution over parameter vectors, because dropping a node is equivalent to setting all the incoming or outgoing weights of that node to zero. With this we can think about the sampling of MC-Dropout as sampling from an unusual probability distribution over weights. Due to the training process, each of these samples is optimized to be as-likely-as-possible. When we then make predictions with MC-Dropout, it is approximately sampling from $p(\theta| D)$. The predictions that they make are samples from the predictive distribution in Equation \ref{eq:aleatoric_epistemic}.

A commonly considered advantage of MC-Dropout is the simplicity with which it can be applied to a Deep Learning model. Many Deep Learning architectures are already trained with dropout, so MC-Dropout can easily be applied without even re-training the model. The big disadvantage however is that it takes many forward passes\footnote{$T = 50$ is recommended, but anywhere from $T=10$ to $T=1001$ may be used. \cite{gal2016dropout, xia2023benchmarking, harper_bayesian_2022}} for the MC-Dropout to capture the predictive distribution, making inference computationally expensive. 

MC-Dropout is therefore easy to apply to architectures that have dropout that have been proven effective, but at the cost of added inference cost. For ECG or EEG monitoring the 100-fold increase in inference cost can be prohibitive.

\subsubsection{Deep Ensembles}

Although there are many ways to do ensembling in Machine Learning, the idea of a Deep Ensemble as an approximation of a Bayesian Neural Network takes the form of several independently trained Neural Networks following the same architecture and trained on the same data~\cite{lakshminarayanan2017simple}\footnote{Originally Deep Ensembles were introduced as a non-Bayesian method for UQ \cite{lakshminarayanan2017simple}, but it has since been shown that it can be considered as a very coarse approximation of a BNN \cite{pearce2020uncertainty, wilson2020bayesian}.}. 

A Deep Ensemble may be interpreted as a small set of samples of the parameter distribution $p(\theta| D)$\cite{jospin2022hands}. Each of these samples is trained to the data, so each sample should reflect a parameter vector with high posterior probability. %

Remarkably, with only a limited number of models\footnote{As an example: \citet{lakshminarayanan2017simple} uses an ensemble of 5 models.} we can achieve an acceptable approximation of the weight distribution $p(\theta | D)$. This keeps the inference cost cheap compared to MC-Dropout, but performing the training several times and storing several models in memory may be expensive. Particularly for applications with model personalization such as in EEG-based BCIs the additional training time can be prohibitive \cite{fawden2023uncertainty}.

Much like MC-Dropout, ensembles are conceptually simple, and intuitive to reason about. It aligns with human analogies where when all the models/people disagree, then there is a lot of (epistemic) uncertainty. Contrastingly, situations where all models/people agree must be very certain.

\citet{xia2023benchmarking} shows that ensembles represent epistemic uncertainty under distributional shifts better than MC-Dropout, and that the accuracy of the predictions is also better. They do this on various Biosignal classification tasks such as auditory COVID-19 classification, respiratory abnormality detection and heart arrhythmia detection. By providing various forms of dataset shift, they concur with findings from computer vision and language models \cite{ovadia2019can}, suggesting that Deep Ensembles may be better at presenting epistemic uncertainty under dataset shifts. In Computer Vision, Deep Ensembles are considered to have state-of-the-art performance for a wide range of Uncertainty Quantification tasks \cite{mucsanyi2024benchmarking}, and from the results of \citet{xia2023benchmarking} it is reasonable to expect that this extends to biosignal applications.

For biosignal applications that do not use Deep Learning alternative ensembling strategies are needed to ensure diversity. \citet{larsen2023new} uses pseudo-bootstrapped \cite{heskes1996practical} ensembles of a logistic regression classifier. In this strategy each model is trained on a subset of the training data to maintain a spread of plausible models. Pseudo-bootstrapping is a viable alternative to achieve ensembling for biosignal applications that use linear classifiers.

\subsubsection{Variational Inference}
In variational inference (VI) the intractable posterior distribution $p(\theta | X, Y)$ is approximated with a simpler distribution $q_\omega(\theta)$. %
A possible approximation through $q_\omega(\theta)$ might say that each weight is a Gaussian distribution with a mean and a variance. The goal is then to optimize the parameters $\omega$ for the high-dimensional Gaussian, so that it is similar to the true posterior. With this, we can then sample models from $q_\omega(\theta)$ to predict class probabilities according to Equation \ref{eq:aleatoric_epistemic}.

In order to make a good approximation of the posterior, VI needs to minimize the Kullback-Leibler divergence (KL-divergence) between the approximate distribution $q_\omega(\theta)$ and the true distribution $p(\theta | X, Y)$. The KL-divergence measures the distance between two distributions. In this case it is given as
\begin{equation}
KL(q_\omega(\theta)\,|| \,p(\theta | X, Y)) = \int_\Theta q_\omega(\theta)\log \frac{q_\omega(\theta)}{p(\theta |\, X, Y)}d\theta.
\end{equation} %
This minimization task still contains the posterior distribution term $p(\theta|X, Y)$ which is intractable as discussed in Equation \ref{eq:bayes_rule}. By rearranging the KL-divergence into the evidence lower bound (ELBO) we instead get the maximization task \cite{abdar2021review}:
\begin{equation}\label{eq:loss_vi}
    \text{ELBO}(\omega) := \int_\Theta q_\omega(\theta) \log p(Y|X, \theta) d\theta - KL(q_\omega(\theta) \,||\, p(\theta))
\end{equation}
The prior chosen for $p(\theta)$ may still be defined by the modeller, and can have an impact on the quality of the model. For the purposes of transfer learning, this prior may even be a learned distribution on another dataset (see \cite{shwartz2022pre}).

While Variational Inference is a more principled approximation of Bayesian Neural Networks than a Deep Ensemble (approximating with a parametric distribution, instead of 5-10 samples), it is also typically less stable in training due to exploding variances. Moreover, implementing it introduces many new decisions to make. The form of the posterior approximation needs to be chosen, as well as the prior for its parameters. Measuring the evidence lower bound requires Monte-Carlo sampling from the approximated posterior. The number of samples is another decision balancing between computational cost per epoch, and the stochasticity of the gradient descent. 

Having many Bayesian layers in a Deep Bayesian Network can cause the loss to become numerically unstable. This instability has made Variational Inference less popular in Computer Vision as they use very large models, but it is not such a big problem for biosignal applications due to the smaller models.

\subsubsection{MCMC sampling of Bayesian Neural Networks}
The most popular methods for Bayesian Neural Networks rely on various rough approximations that give empirically good results, because it is generally considered infeasible to train large BNNs otherwise. 

A theoretically ideal approach to Bayesian Neural Networks would sample the weights according to the posterior $\theta \sim p(\theta|X,Y)$. This can be done with the Metropolis-Hastings algorithm which proposes many possible weights $\theta$, and accepts them or not based on the posterior. By sampling arbitrarily many weight configurations, we get an arbitrarily good approximation of the Bayesian Neural Network. Inference is then done by making a prediction with each sampled weight configuration. 

It is clear that for very large BNNs with very large datasets this is computational infeasible, but specifically Machine Learning on biosignals typically uses comparatively small Neural Networks with comparatively small datasets. This makes MCMC-based BNN approximations ideal for biosignal applications. 

We found only one studying implementing a form of MCMC-based BNN approximations. \citet{chetkin2023bayesian} used Adaptive Stochastic Gradient Hamiltonian Monte Carlo \cite{chen2014stochastic} for Motor Imagery classification. They found that this worked better than an ensemble when applied to ShallowConvNet \cite{schirrmeister2017deep}, but there was no statistically significant difference when applied to EEGNet \cite{lawhern2018eegnet}. \citet{zhang2023neuromusculoskeletal} uses it for knee torque regression based on EMG and finds it gives comparable prediction and uncertainty estimation to Gaussian Processes. 

Future work exploring MCMC approximations for BNNs can give improved uncertainty estimates, especially for small datasets. Even an increased latency for a diagnosis of hours or even a few days may be warranted if it improves uncertainty estimates and subsequently clinical outcomes. We recommend \citet{chandra2024bayesian} for a tutorial to encourage future efforts in this direction.

\subsection{Evidential Deep Learning}\label{sec:edl}

Evidential Deep Learning \cite{sensoy2018evidential} offers a computationally affordable alternative to Bayesian Neural Networks where the distribution of the predictions is captured by a $c$ dimensional Dirichlet distribution parameterised by $\alpha_c \in [0, 1]$, which are predicted by a Neural Network. This setup therefore predicts a distribution of probabilities in a single forward pass. 

\citet{sensoy2018evidential} proposed EDL to look at uncertainty from the perspective of the Dempster-Shafer Theory of Evidence (DST) instead of the aleatoric-epistemic approach. In this approach the parameter $\alpha_c$ gives the amount of \textit{evidence} for that class. 

The uncertainty can then be defined into \textit{vacuity} and \textit{dissonance}~\cite{lin_reliability_2022, lin_robust_2023}. Vacuity is the absence of evidence causing uncertainty. Like standard Neural Networks with a Softmax activation function, Evidential Machine Learning assumes that exactly one class must be the ground truth. The absence of evidence for any of the classes would then result in a form of uncertainty referred to as vacuity. The opposite uncertainty is \textit{dissonance}, which occurs when the model has found evidence for multiple classes, which is not in line with the assumption of mutual exclusivity.

Prior Networks \citep{malinin2018predictive} are another approach form of Evidential Deep Learning. It uses the same setup of predicting a Dirichlet distribution but interprets it as an alteration of the aleatoric-epistemic uncertainty. Under the Bayesian Neural Network framework we consider the uncertainty due to generalization error, such as when the model is evaluated under out-of-distribution data, as part of the epistemic uncertainty. Prior networks treat the mismatch between train and test distributions separately under the term \textit{distributional uncertainty}. This then gives 
\begin{equation}
p_c = \int \int \underbrace{p(y\,{=}\,c\,|\,\mu)}_{\text{aleatoric}}   \underbrace{p(\mu\,|\,x, \theta)}_{\text{distributional}} \underbrace{p(\theta\,|\,D)}_{\text{epistemic}} d\mu \,d\theta.
\end{equation}

Evidential Deep Learning has shown good performance on out-of-distribution detection tasks, but has theoretical and practical limitations when representing epistemic uncertainty \cite{jurgens2024epistemic}. The reviewed literature generally does not compare EDL methods with BNN methods for biosignal applications. A thorough investigation of uncertainty quantification should consider both top-performing methods for BNNs and EDL methods for biosignal tasks. From the current literature, it can only be established that EDL methods give better uncertainty estimates than standard Neural Networks \cite{lin_reliability_2022, li_real-time_2022} on EMG grasp classification and that its uncertainty indeed goes up with noise for myocardial infarction detection under noisy ECG \cite{jahmunah_uncertainty_2023}.

\subsection{Gaussian Process Regression}
Gaussian Process Regression \cite{schulz2018tutorial} is a non-parametric regression method that considers epistemic uncertainty. It assumes a Gaussian prior over the dependent variable $Y$. It also assumes that the samples in the training data $D$ are drawn with a known amount of measurement error (aleatoric uncertainty). This leaves epistemic uncertainty in the regression between and outside training samples, and gives more certainty at points close to the training samples. 

As more training samples are collected, the epistemic uncertainty will decrease. The aleatoric uncertainty is fixed (either heteroscedastically or homoscedastically). The amount of aleatoric uncertainty needs to be known for each datapoint, or assumed, but cannot be learned from the data. 

Gaussian Process Regression is suitable for biosignal applications due to the smaller datasets, but because most tasks are classification tasks it does not see a lot of use. Current implementations on EMG \cite{zhang2023knee} and ECG \cite{costabal_machine_2019} demonstrate it as an effective method in combination with physics-informed simulation models. %

\tabcolsep 9pt
\renewcommand\arraystretch{1.3}
\begin{table*}[ht]
\centering
\caption{\label{3} A simpified overview of the different UQ methods discussed in \ref{sec:UQ-Methods}. Computational cost of UQ methods is qualitatively grouped into 3 classes. \textit{None} has negligible added computational cost. \textit{Small} has some added computational cost e.g. due to slower convergence or training steps being more computationally expensive. \textit{Large} indicates substantial increase in computational cost, such as 5 times the training cost, or 50 times the inference cost.}\vspace{1mm}
{%
\adjustbox{max width=\linewidth}{%
 \begin{tabular}{ll l l l l}
        \toprule
         Method & Model Agnostic & Epistemic UQ & Aleatoric UQ** & Training Cost & Inference Cost \\
         \midrule
         MC-Dropout \cite{gal2016dropout} & NN only & \checkmark & & None & Large \\
         Ensembles \cite{lakshminarayanan2017simple}& \checkmark* & \checkmark & & Large & Small \\
         Variational Inference \cite{hoffman2013stochastic} & NN only & \checkmark & & Large & Large\\
         Variational Autoencoder \cite{kingma2019introduction} & & & \checkmark & Small & Large \\
         Evidential Machine Learning \cite{sensoy2018evidential, malinin2018predictive} & \checkmark & \checkmark*** & \checkmark & None & None \\
         Gaussian Process Regression \cite{williams2006gaussian} & & \checkmark & \checkmark & Small & Small \\
         Post-hoc calibration \cite{guo2017calibration}& \checkmark & & \checkmark & None & None \\

        \bottomrule
         
\end{tabular}}%
\begin{flushleft}
\footnotesize
*Requires bootstrapping \cite{heskes1996practical} for non-stochastic training procedures. May perform poorly without local minima.\\
**Aleatoric uncertainty may still show in classification with Softmax.\\
***Not faithful epistemic uncertainty \cite{jurgens2024epistemic}.

\end{flushleft}}
\label{tab:uq_methods_overview}
\end{table*}

\subsection{Post-hoc Calibration}

Post-hoc calibration methods~\cite{guo2017calibration} look at uncertainty only in terms of the predicted probability for each class, and addresses how this may deviate from the observed probability. A class prediction with $p=0.75$ should be correct $75\%$ of the time, but this does not hold for standard softmax classification. %
Post-hoc calibration methods aim for an optimal calibration such that 
\begin{equation}
    p(y\,{=}\,c \,|\, p_c) = p_c.
\end{equation}
Various methods for post-hoc probability calibration methods exist \cite{guo2017calibration}. Temperature Scaling is the simplest method of post-hoc calibration, which determines the \textit{softness} of the Softmax function. It does so by introducing a hyperparameter $\tau$ to get the scaled Softmax function
\begin{equation}\label{eq:softmax_temerature}
    p(y\,{=}\,c\,|\, x, \theta) = \frac{\exp(\frac{f_c^{\theta}(x)}{\tau})}{\sum_c' \exp(\frac{f_{c'}^{\theta}(x)}{\tau})}.
\end{equation}
Post-hoc calibration methods cannot provide better separation between correct and incorrect predictions, and do not account for epistemic uncertainty. They only ensure that the probabilities are appropriately scaled, which is important when those probabilities need to be interpreted by a clinician \cite{elul_meeting_2021}.

\subsection{Non-standard UQ Methods}
Above, a selection of common and well-studied methods for Uncertainty Quantification is discussed. This does not cover all the UQ methods that were encountered in the review. Below we continue the description of uncertainty quantification methods with some non-standard methods encountered in the reviewed literature to provide an exhaustive presentation of UQ research on biosignals. 

Biosignal classification often uses smaller models and smaller datasets than computer vision, which makes it suitable for unique Uncertainty Quantification methods that are not standard in other domains. We critically assess these methods below.

\subsubsection{Early Exit Ensembles}
\begin{figure}[t]
    \centering
    \includegraphics[width=\columnwidth]{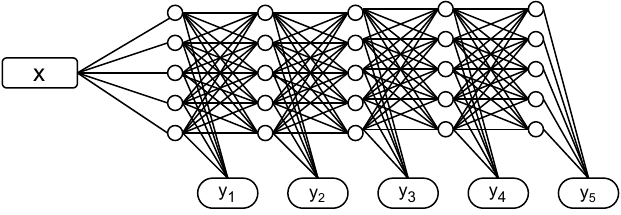}
    \caption{A diagram showing the concept of Early Exit Ensembles \cite{campbell_robust_2022}. There is a shared backbone network, from which Exit Branches make predictions. Each exit branch makes an independent prediction. The distribution of predictions may be interpreted similar to a normal ensemble.}
    \label{fig:early_exit_ensemble}
\end{figure}

As a quasi-ensembling method \citet{campbell_robust_2022} propose Early Exit Ensembles. Early exit ensembles work by taking any deep neural network and adding various \textit{exit} branches to points of the network as illustrated in Figure \ref{fig:early_exit_ensemble}. %
The idea is that each exit branch will try to learn to do the classification task (as an ensemble), but depending on the location on the \textit{backbone} architecture they may learn on either lower or higher level features.

Like normal ensembling methods, the disagreement between the various classifiers corresponds to epistemic uncertainty. The weight sharing reduces the computational cost of training and inference, making it more suitable for edge devices \citet{fawden2023uncertainty}. The ways in which constructing an Early Exit Ensemble from an existing architecture affects the quality of the predicted uncertainty is an interesting avenue for research, which may be partly inspired by what is already known about \mbox{early-exit neural networks (see \cite{teerapittayanon2016branchynet, montanari2020eperceptive}).} %

\subsubsection{Bayesian Model Averaging with Reversible-Jump MCMC}
\citet{schetinin_bayesian_2007} attempted to classify EEG artifacts using a method based on Bayesian Model Averaging. They sample changes to a decision tree, that are either accepted or rejected based on the likelihood given the data. %

As a measure of uncertainty the authors consider the entropy in the leaf nodes. %
Since the dataset is not specified and no other models are shown it is not possible to assess the quality of the model, nor the resulting entropies. Another reviewed work also used the entropy of the leaf-nodes in a decision tree as a measure of uncertainty, but this also lacked interpretation \cite{hagan_comparison_2021} .

\subsubsection{Majorization-Minimization and Hierarchical Bayesian Modelling} \label{sec:mm_hbm}
\citet{bekhti_hierarchical_2018} compares Majorization-Minimization and Hierarchical Bayesian Modelling and shows how they are fundamentally the same. While other works aim to learn a function $p(y|x, D) = f_\theta(x)$, this work starts with the assumption that observed EEG recordings represented as a matrix $M$ are a linear combination of underlying sources $X$ connected through a known linear forward propagation matrix $G$, with some Gaussian noise $E$ such that $M = GX + E$. This results in a multi-task regression where we need to learn an optimal matrix $X$ to find the underlying sources. This results in the optimization problem
\begin{equation}
    \hat{X} = \argmin_X \frac{1}{2}||M-GX||_F^2.
\end{equation}

Using Majorization-minimization the authors generate multiple sparse solutions, together with how well they minimize the objective function. This offers various source attributions to an observed EEG or MEG signal, together with a measure of how (un)certain each solution is, giving a more complete insight into the source of a given signal.

\subsection{Variational Autoencoders}\label{sec:VAE}
A Variational Autoencoder \cite{kingma2019introduction} is a specific type of neural network architecture. It has an encoder which receives a high-dimensional input $x$ and encodes it into a lower dimensional latent distribution $p(z\,|\,x, \theta)$. It does so by predicting a mean and a variance for each dimension of the latent distribution, from which latent representations $z \sim p(z\,|\,x)$ can be sampled. A decoder network then reconstructs the encoding back into the original dimensionality of the input to achieve $x' = f_{\theta'}(z \sim p(z|x, \theta)) \approx x$. 

The VAE model is trained to minimize the difference between the input $x$ and the reconstructed output $x'$. As a result, the latent distribution $p(z\,|\,x, \theta')$ should be a lower-level representation of the salient features that exist in the data. %
Because the latent representation is a distribution which can be sampled from, researchers have constructed various methods to extract uncertainty from that stochasticity. \citet{belen_uncertainty_2020} uses a trained VAE on a dataset of segments of ECG with and without expert annotated atrial fibrillation. They then use the sampled latent representations as input for a multi-layer-perceptron to do the classification task as
\begin{equation}
    p(y=c\,|\,p(z|x, \theta), \theta').
\end{equation}
This results in a distribution of probabilities, of which the variance is used to measure aleatoric uncertainty. 

\citet{van_de_leur_interpretable_2021} apply Principal Component Analysis to get a 2-dimensional visualization of the latent space as a method for interpretability for ECG classification. They show how various diagnoses would show in the latent representation, so that a sample on the boundary of two classes, or far away from any known classes can be qualitatively assessed as uncertain. This shows unique opportunities for using VAEs for uncertainty. 

The primary downside to using VAEs for Biosignal analysis is that it imposes specific architecture constraints. A lot of the biosignal literature relies on established architectures that are known to perform well in similar tasks, but those cannot easily be turned into VAEs. %

\subsubsection{Assumed Density Filtering} \label{sec:ADF}
\citet{duan_uncer_2023} applies a computationally affordable method for modelling data uncertainty called Assumed Density Filtering (ADF). Whereas Bayesian Neural Networks model a distribution for each weight, ADF takes a single-point solution for the weights, but has a distribution for the activations, by modelling the input as a Gaussian distribution around the single-point input features. %

Each activation is modelled by a mean and variance, where the variance corresponds to the uncertainty. This method is intended to correspond to aleatoric uncertainty caused by uncertain inputs. For biosignals this may represent sensor noise or artifacts. Combined with a Bayesian Neural Network as done by \citet{duan_uncer_2023} provides explicit modelling for both uncertainty of the model, and uncertainty of the biosignal recording. %

They demonstrate that this gives better uncertainty estimates for a BCI task than many alternative methods including Deep Ensembles \cite{lakshminarayanan2017simple}, MC-Dropout \cite{gal2016dropout} and Direct Uncertainty Quantification \cite{van2020uncertainty}. %

\subsubsection{Data Uncertainty Learning}

As a method for aleatoric uncertainty, Data Uncertainty Learning \cite{chang2020data} models uncertainty as a distribution in an embedding such that
\begin{equation}
    p(z|x) = \mathcal{N}(x; \mu, \sigma^2 I). 
\end{equation}

This is similar to a Variational Autoencoder, but where autoencoders have the embedding as the middle layer, Data Uncertainty Learning has the embedding as the penultimate layer. The decoder is then replaced with a shallow classifier. 

\citet{deng_eeg-based_2023} applied this method to predict seizures from EEG. The uncertainty in the embedding should then capture the uncertainty that is in the EEG recording. Since the uncertainty is modelled in a deep embedding it may represent more nuanced uncertainty in the EEG signal that relates directly to the task. \citet{deng_eeg-based_2023} show that the modelling of uncertainty improves the classifier as compared to a deterministic equivalent. They find that wrong predictions indeed are more likely to have high uncertainty, but a thorough evaluation and comparison with alternative methods is still needed.

\subsubsection{Neural Stochastic Differential Equations}
\citet{wabina_neural_2022} propose a novel method called Neural Stochastic Differential Equations (SDE-Net) to learn an electrical conductivity model of the head from MRI. Such conductivity models can be used to inform the forward propagation of EEG signals as mentioned to in Section \ref{sec:mm_hbm}. 

They use a class of Deep Neural Networks proposed in \cite{kong2020sde}, which includes a split block consisting of a drift and a diffusion network to consider the Neural Network as a Stochastic Differential Equation. The drift network continues to attempt to optimize predictions, while the diffusion network predicts heteroscedastic Gaussian noise. The noise should be minimal for samples in the training distribution, and maximal for out-of-distribution samples. %

An experiment on the Single Individual volunteer for Multiple Observations across Networks \mbox{(SIMON)} MRI dataset showed that SDE-Net outperformed Bayesian methods. However, the effect of epistemic uncertainty on the spread of the predictions and SDE-Net's ability to capture epistemic uncertainty is not investigated, so the results may be explained by better estimation of aleatoric uncertainty alone.

\subsubsection{Reconstruction Error}
\citet{martinez_strategic_2020} look at how to reconstruct an ECG signal based on bioimpedance recordings. Bioimpedance can be much easier to record, but also difficult for cardiologists to interpret. They propose a method where an Autoencoder regression model uses the biosignals to construct the ECG morphology, but without correct amplitudes. Then a second autoencoder uses this amplitude-invariant data, and the original bioimpedance to reconstruct the ECG. 

Differences in morphology are attributed to a generalisation failure of the second autoencoder. Thus, the authors measure the Pearson correlation between the amplitude invariant and amplitude-corrected data as a measure of uncertainty. 
They show that this uncertainty indeed correlates with the translation quality, but a thorough comparison with more established UQ methods is still needed. 

\subsubsection{Fuzzy Logic}
The systematic search found three works that rely on methods from Fuzzy Logic. Fuzzy Logic relies on the concept of a Fuzzy Set with a fuzzy membership function. This gives a probabilistic notion of a set where an element can have partial membership to a set or multiple sets. 

The fuzzy membership function may be defined in different ways, based on (fuzzy) unsupervised clustering  \cite{sovatzidi_constructive_2022}, (fuzzy) classification \cite{liu2017weighted} or as a composite of other fuzzy membership functions \cite{mishra2023cardiolabelnet}. 
The fuzzy memberships can be interpreted directly as predictions \cite{mishra2023cardiolabelnet}, but more complex setups are also possible. \citet{sovatzidi_constructive_2022} uses them to construct a Fuzzy Cognitive Map: A directed probabilistic graph that offers an explainable decision support system \cite{amirkhani2017review} for diagnosis. \citet{liu2017weighted} instead uses it to combine predictions from multiple modalities using the Dempster-Shafer Theory of Evidence. They show that their proposed Weighted Fuzzy Dempster-Shafer Framework can fuse predictions from different modalities to achieve better predictive performance than either modality alone. 

Fuzzy Logic allows a lot of freedom for the modeller to design probabilistic systems, which is relevant for biosignal analysis where we have limited datasets but do have prior knowledge on how a decision should be made. We find that the proposed works have good reason for their design and show improved task performance, but a systematic evaluation of predictive uncertainty under such Fuzzy Logic systems is still missing.

\subsubsection{Bayesian Moderated Outputs}
Based on \citet{mackay1992bayesian}, \citet{mohamed_detection_2005} compare Bayesian Moderated Outputs to a standard Multi-Layer Perceptron for the task of epileptic activity classification in sleep EEG recordings. The concept of Bayesian Moderated Outputs is that instead of having a single optimal parameter vector $\hat{\theta}$, a more robust method will have a Gaussian distribution of parameters around an optimum $\Theta = \mathcal{N}(\hat{\theta}, s^2)$. %

Unfortunately, this did not lead to apparent better performance than a maximum-likelihood trained Multi-Layer Perceptron \cite{mohamed_detection_2005}. %
The Bayesian Moderated Outputs did achieve slightly higher accuracy (up to 1 percent-point), but at the cost of rejecting up to 15 percent-point more samples from classification.

\begin{table*}[!ht]
    \centering
            \caption{Reviewed papers and their Uncertainty Quantification methods. }\vspace{0.5em}
    \label{tab:overview_papers}
    \adjustbox{max width=\linewidth}{%

    \begin{tabular}{l p{0.2\linewidth}   p{0.2\linewidth}   p{0.2\linewidth}}
            \toprule
    
         \textbf{UQ Method} & \textbf{EEG publications} & \textbf{ECG publications} & \textbf{Other biosignal}  \\\midrule
         \textbf{MC-Dropout \cite{gal2016dropout}} & Epilepsy \cite{borovac_calibration_2022, wong2023estimating, campbell_robust_2022}, Sleep \cite{fiorillo_deepsleepnet-lite_2021}, Motor Imagery BCI \cite{milanes-hermosilla_monte_2021, duan_uncer_2023}, Denoising \cite{jin2023uncertainty} & Emotion \cite{harper_bayesian_2022}, Respiration \cite{rathore_multifunctional_2023}, Arrhythmia \cite{xia2023benchmarking, barandas_evaluation_2024, elul_meeting_2021, aseeri_uncertainty-aware_2021, vranken2021uncertainty, islam2022monte, mendoza2023deep, zhang2024cardiac}, Anxiety \cite{zanna_bias_2022} & EOG: Ataxia \cite{stoean_automated_2020}, MRI: Focal Cortical Dysplasia \cite{gill_multicenter_2021} \\\midrule
         
         \textbf{(Deep) Ensemble \cite{lakshminarayanan2017simple}} & Motor BCI \cite{duan_uncer_2023} & Arrhythmia \cite{xia2023benchmarking, strodthoff_deep_2021, barandas_evaluation_2024, park_self-attention_2023, aseeri_uncertainty-aware_2021, vranken2021uncertainty}, CRT response \cite{larsen2023new} & \\\midrule
         
         \textbf{Variational Inference \cite{gal2015bayesian}} & P300 BCI \cite{ma_bayesian_2023}, Motor BCI \cite{milanes-hermosilla_robust_2023} & Arrhythmia \cite{xia2023benchmarking, vranken2021uncertainty, rahman2023quantifying}, & fNIRS: Motor BCI \cite{siddique2021classification} \\\midrule

         \textbf{Softmax \cite{bridle1990probabilistic}} & Sleep \cite{phan_sleeptransformer_2022}, Epilepsy \cite{vavaroutas2023uncertainty} & Arrhythmia \cite{xia2023benchmarking, vavaroutas2023uncertainty} & \\\midrule

         \textbf{Variational Autoencoders \cite{kingma2019introduction}} & &  Arrhythmia \cite{xia2023benchmarking, van_de_leur_interpretable_2021, belen_uncertainty_2020}, Modality Translation \cite{martinez_strategic_2020} & \\\midrule
         
         \textbf{Evidential Deep Learning \cite{sensoy2018evidential}} & & Myocardial Infarction \cite{jahmunah_uncertainty_2023} & EMG: Hand movement \cite{lin_reliability_2022, lin_robust_2023}\\\midrule

          \textbf{Post-hoc calibration \cite{guo2017calibration}} & Sleep \cite{fiorillo_deepsleepnet-lite_2021} & Arrhythmia \cite{xia2023benchmarking, barandas_evaluation_2024} & \\\midrule
         
         \textbf{Gaussian Process \cite{schulz2018tutorial}} & Motor BCI \cite{duan_uncer_2023} & Heart Modelling \cite{costabal_machine_2019} & EMG: Knee torque \cite{zhang2023knee} \\\midrule
         
         \textbf{Heteroscedastic UQ \cite{seitzer2022pitfalls}} & Motor BCI \cite{duan_uncer_2023}, Denoising \cite{jin2023uncertainty} & Arrhythmia \cite{vranken2021uncertainty} & \\\midrule

         \textbf{Early Exit Ensemble \cite{campbell_robust_2022}} & Epilepsy \cite{campbell_robust_2022, fawden2023uncertainty} & & \\\midrule

         \textbf{Hamiltonian Monte Carlo \cite{chen2014stochastic}} & Motor BCI \cite{chetkin2023bayesian} & & EMG: Knee torque \cite{zhang2023neuromusculoskeletal} \\\midrule

         \textbf{Fuzzy Sets \cite{amirkhani2017review}} & Depression \cite{sovatzidi_constructive_2022} & Arrhythmia \cite{mishra2023cardiolabelnet} & \\\midrule

         \textbf{Bayesian Model Averaging \cite{fragoso2018bayesian}} & Sleep \cite{schetinin_bayesian_2007} & & \\\midrule
         \textbf{Hierarchical Bayesian Modelling} & Inverse Problem \cite{bekhti_hierarchical_2018} & & \\\midrule
         \textbf{Entropy in Decision Tree Leafs} && Arrhythmia \cite{hagan_comparison_2021} & \\\midrule
         \textbf{Bayesian Moderated Output \cite{mackay1992bayesian}} & Epilepsy \cite{mohamed_detection_2005} && \\\midrule
         \textbf{Direct Uncertainty Learning \cite{chang2020data}} & Epilepsy \cite{deng_eeg-based_2023} && \\\midrule
         \textbf{Kalman Filters \cite{mandic2015intrinsic}} & Epilepsy \cite{de_rooij_enabling_2023} && \\\midrule
         \textbf{Deep SVDD \cite{ruff2018deep}} & Epilepsy \cite{wong2023estimating} && \\\midrule
         \textbf{Neural SDE \cite{kong2020sde}} & & & MRI: Inverse Problem \cite{wabina_neural_2022} \\\midrule
         \textbf{Assumed Density Filtering \cite{gast2018lightweight}} & Motor BCI \cite{duan_uncer_2023} & & \\\midrule
         \textbf{DUQ \cite{van2020uncertainty}} & Motor BCI \cite{duan_uncer_2023} && \\\midrule
         \textbf{WFDSG \cite{liu2017weighted}} & Drowsiness \cite{liu2017weighted}& & EOG: Drowsiness \cite{liu2017weighted} \\\midrule
         \textbf{Trust Scores \cite{jiang2018trust}} && Arrhythmia \cite{li2023effect} & \\

        \bottomrule

            \end{tabular}
            }

\end{table*}

\begin{figure*}[b]
\centering
\subfigure[Low Aleatoric \& Low Epistemic]{
  \begin{minipage}[t]{0.21\linewidth}
  \centering
  \includegraphics[width=\linewidth]{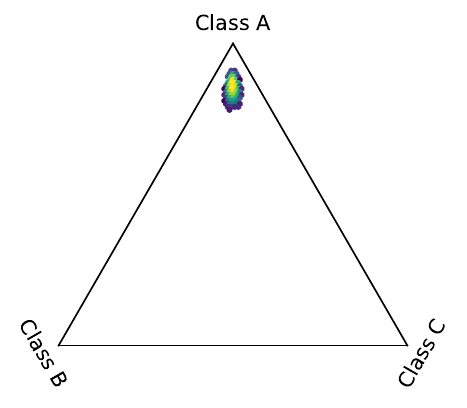}
  \end{minipage}
}
\subfigure[Low Aleatoric \& High Epistemic]{
  \begin{minipage}[t]{0.21\linewidth}
  \centering
  \includegraphics[width=\linewidth]{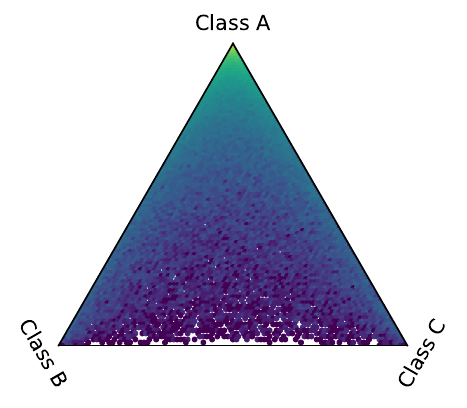}
  \end{minipage}
}
\subfigure[High Aleatoric \& Low Epistemic]{
  \begin{minipage}[t]{0.21\linewidth}
  \centering
  \includegraphics[width=\linewidth]{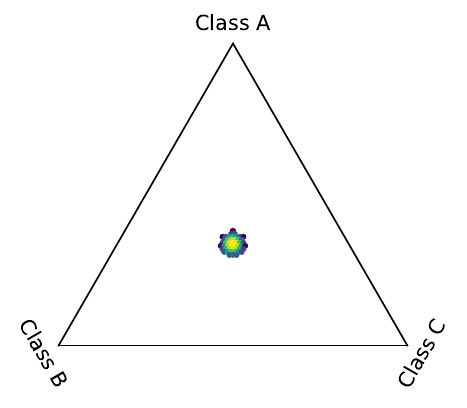}
  \end{minipage}
}
\subfigure[High Aleatoric \& High Epistemic]{
  \begin{minipage}[t]{0.265\linewidth}
  \centering
  \includegraphics[width=\linewidth]{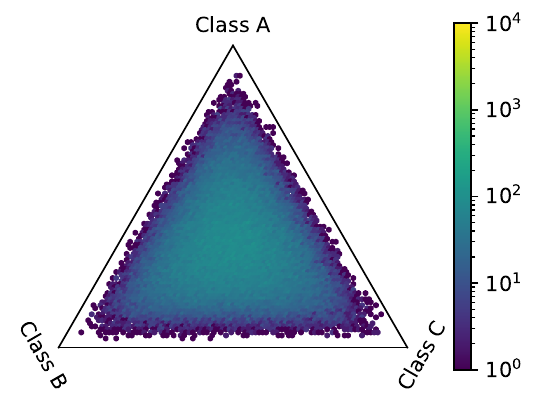}
  \end{minipage}
}
\caption{Simplexes presenting different types of uncertainty. Epistemic uncertainty is presented by increased variance in the logits. Aleatoric uncertainty is presented by decreasing the difference between the means of the logits between classes. The points
represent softmax probabilities determined by logits following a multivariate Gaussian $\mathcal{N}(\mu, \sigma^2)$, with colour representing frequency. For high aleatoric uncertainty we set $\mu = [10, 10, 10]$, whereas for low we use $\mu = [10, 8, 8]$. For high epistemic uncertainty we set $\sigma^2 = [2, 2, 2]$, whereas for low we use $\sigma^2 = [0.01, 0.01, 0.01]$.}
\label{fig:simplex}
\end{figure*}

\subsubsection{Miscellaneous methods}
We encountered two more uncertainty quantification methods, but they were sufficiently rare that they do not fit into the presented narrative. 

To deal with the large amount of data in the Temple University Hospital Seizure Corpus (TUSZ) \cite{obeid2016temple} dataset, \citet{de_rooij_enabling_2023} used Kalman Filters to solve the least squares adaption of SVMs. Rather than optimizing the SVM for epilepsy classification against the whole dataset at once, they consider parts of the dataset to continually learn the parameters of the SVM. Since Kalman Filters allow for some uncertainty, this method should capture model uncertainty. However, the authors do not go into detail on how well the uncertainty quantification performs.

Other work from \citet{li2023effect} investigated Trust Scores \cite{jiang2018trust} under dimensionality reduction. Trust Scores assign uncertainty based on disagreement between a proposed model and a kNN-based classifier, where high disagreement indicates high uncertainty. However, they found that for some dimensionality reduction methods the uncertainty was not monotonically increasing with the precision, indicating a potential risk when implementing Trust Scores in a classification pipeline.

\subsection{Recommendations for UQ methods}

We conclude from the analysis of UQ methods that Deep Ensembles and MC-Dropout are the best established, and that Deep Ensembles may be considered state-of-the-art for estimating epistemic uncertainty. This is also supported by findings from \citet{mucsanyi2024benchmarking}, which shows that Deep Ensembles performs well in many contexts. The review found relatively little comparative analysis, specifically we find that comparing a computationally expensive Bayesian Neural Network against a standard single-point Neural Network for uncertainty estimation is necessary for all implementations. The class probabilities from standard Neural Networks are often already quite good estimates of uncertainty \cite{manivannan2024uncertainty, suurmeijer2025uncertainty}.

Post-hoc calibration gets little attention in the presented research, but is valuable for addressing overconfidence \citet{suurmeijer2025uncertainty} and ensuring clinically interpretable predictive probabilities. We encourage future work to combine Deep Ensembles with post-hoc calibration as de-facto method for achieving good uncertainty estimates in classification, and heteroscedastic uncertainty models combined with Deep Ensembles for regression. 

This combination yields well-calibrated uncertainty estimates, are somewhat robust to epistemic uncertainty, and are well-suited for detecting out-of-distribution examples \cite{mulder2026challenge}. As far as we are aware, no method has been shown to consistently outperform this approach.

\section{Uncertainty Measures} \label{sec:UQ_measures}

    Most uncertainty quantification methods (e.g. BNNs, EDL) when applied in classification tasks produce a distribution over class probabilities. However, upon reviewing the biosignal literature we found that papers are inconsistent or non-specific about how to extract scalar measures of uncertainty from them\footnote{In regression the literature is more consistent: Variance from either aleatoric or epistemic methods indicate the source of uncertainty, as shown in \citet{jin2023uncertainty}. Whether this separates aleatoric ande epistemic uncertainty correctly in practice is still unknown \cite{mucsanyi2024benchmarking}.}. We critically review the existing \textit{Uncertainty Measures} in relation to theoretical expectations of aleatoric and epistemic uncertainty, and provide strong recommendations on how to extract uncertainty measures from a distribution of predicted probabilities. An overview is given in Table \ref{tab:uq_measures_overview}.

    The theoretical analysis considers whether the measure captures aleatoric uncertainty, epistemic uncertainty or both. It relies on the notion that epistemic uncertainty is represented by disagreement between model predictions. 

    Figure \ref{fig:simplex} shows how aleatoric and epistemic uncertainty interact \cite{de2024disentangled, wimmer2023quantifying}. These plots are generated by taking 3 Gaussian distributions to represent predicted logits. 100.000 samples are taken from these logits and passed through the Softmax function. The closeness to each vertex represents the predicted class probability. This provides an intuition of how aleatoric and epistemic uncertainty may present as predicted class probabilities. It becomes apparent that under high epistemic uncertainty (Figures \ref{fig:simplex}.b, \ref{fig:simplex}.d), determining aleatoric uncertainty becomes difficult as the distribution gets smoothed. Under maximal epistemic uncertainty, all values of aleatoric uncertainty would become equally likely. \citet{wimmer2023quantifying} explores axiomatically how this second-order distribution approach to epistemic uncertainty must cause interactions between the estimates for some measures of uncertainty.  %

    \subsection{Class Probability}
    The standard method for measuring uncertainty in Neural Networks is the predicted Softmax probability of a classification. An epilepsy classifier that gives the diagnosis of epilepsy with $p=0.55$ is less certain than if it gives the diagnosis with $p=0.97$. 
    
    This uncertainty measure captures aleatoric uncertainty. %
    However, softmax probabilities are infamously overconfident in single-point neural networks, even when using a proper scoring loss function \cite{guo2017calibration}.

    When multiple forward passes are made with a BNN the class probability is determined by the average of all forward passes. With $T$ as the number of forward passes and $\bar{c}$ as the max probability class of the average probabilities ($\bar{c} = \argmax_c  T^{-1}\sum_{t} p_{c}$) we define the class probability as:
    \begin{equation}
        \mathbb{P}(p) \equiv T^{-1}\sum_T p_{\bar{c}}
    \end{equation}
    Or in a shorthand:
    \begin{equation}
        \mathbb{P}(p) \equiv \bar{p}_{\bar{c}}
    \end{equation}    
        \begin{figure}[t]
            \centering
            \includegraphics[width=0.8\columnwidth]{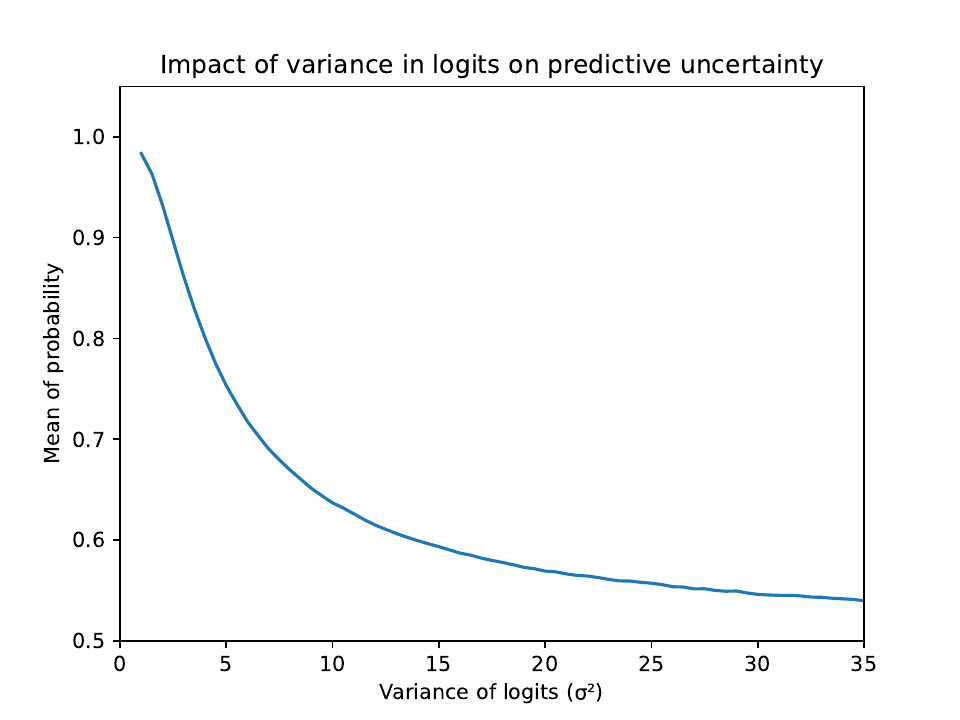}
            \caption{Mean class probability decreases for higher variance in the logits. This illustration assumes binary classification. The logits are distributed as $\mathcal{N}([4, 0], [\sigma^2, \sigma^2])$. Only the first class is shown.}
            \label{fig:mean_logit_variance}
        \end{figure}    
        For approximations of Bayesian Neural Networks we can assume that the logits increase in variance as the epistemic uncertainty increases. The Softmax function pushes high logits down into a $[0, 1]$ range, while lower logits are shifted less. As such, logits from a distribution with high variance will result in less confident probabilities. Figure \ref{fig:mean_logit_variance} visualizes this effect. 

        This shows that the average class probabilities $\mathbb{P}(p)$ will show more uncertainty under increased epistemic uncertainty. Therefore, it is a measure that combines aleatoric and epistemic uncertainty. This explains why the average probability $\mathbb{P}(p)$ of a BNN is less overconfident than a single-point Neural Network \cite{fiorillo_deepsleepnet-lite_2021, aseeri_uncertainty-aware_2021}.

    \subsection{Variance}
    Several papers consider the variance or standard deviations of the class probabilities as a measure of uncertainty \cite{stoean_automated_2020, zanna_bias_2022, fiorillo_deepsleepnet-lite_2021, strodthoff_deep_2021, elul_meeting_2021, harper_bayesian_2022}. This should represent epistemic uncertainty as it measures disagreement between model samples. 

    Under multi-class classification it can be unclear which variance should be computed. Some implementations measure the variance over each class and either present all those variances to clinicians \cite{elul_meeting_2021} or as features to another Machine Learning model \cite{stoean_automated_2020}.  One may also present only the variance of the predicted class as a measure of epistemic uncertainty, or the average variance over multiple classes.

    To be specific, this leaves two possible scalar measures for probability variance under multi-class predictions:

    \begin{figure}[t]
        \centering
        \includegraphics[width=0.8\columnwidth]{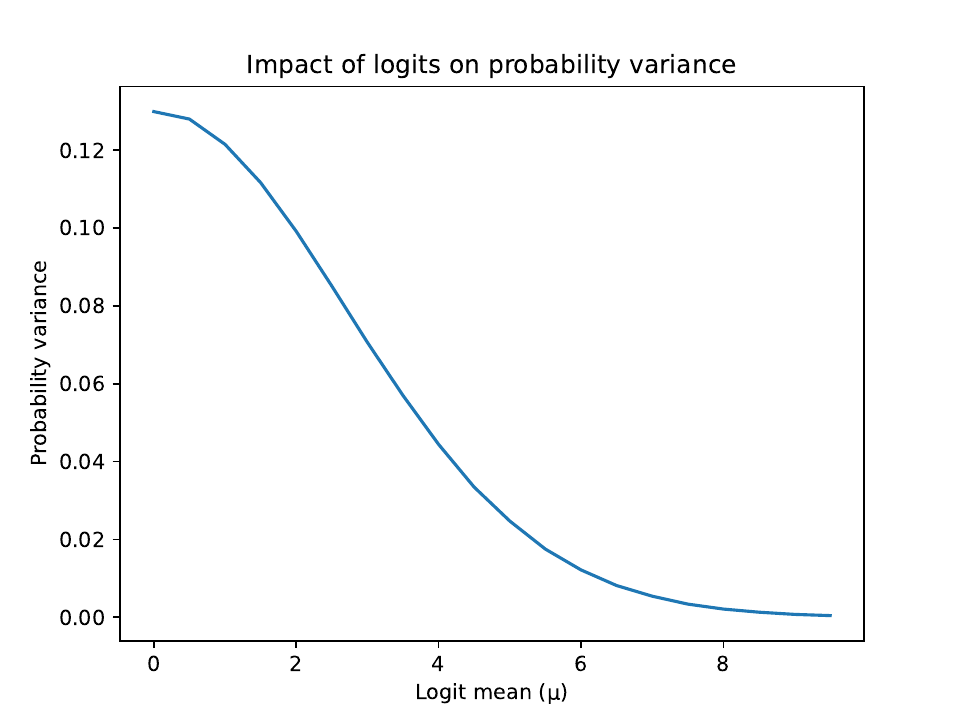}
        \caption{Probability variance decreases as the logit mean increases. This illustration was generated by taking logits as $\mathcal{N}([\mu, 0], [2, 2])$. Here we see that probability variance (used as a measure of epistemic uncertainty) becomes smaller when aleatoric uncertainty decreases.}
        \label{fig:variance_logit_mean}
    \end{figure}
    \begin{equation}\label{eq:class_var_pred}
        \mathbb{V}_{\bar{c}}(p) = T^{-1} \sum_t (p_{\bar{c}t} - \bar{p}_{\bar{c}})^2 
    \end{equation}    
    \begin{equation} \label{eq:class_var_all}
        \mathbb{V}(p)= C^{-1} \sum_c  T^{-1}\sum_t  (p_{ct} - \bar{p}_{c})^2 
    \end{equation}
    A similar effect as shown in Figure \ref{fig:mean_logit_variance} occurs when applying variance uncertainty measures to the class probabilities. Figure \ref{fig:variance_logit_mean} illustrates that the difference in the mean of the logits increases (less aleatoric uncertainty) the variance of the class probabilities decreases. As a result a decrease in aleatoric uncertainty can present as a perceived decrease in epistemic uncertainty. Future works should consider using the variance of the logits as described in \cite{valdenegro2022deeper} to get a more independent measure of epistemic uncertainty.

\tabcolsep 9pt
\renewcommand\arraystretch{1.3}
\begin{table*}[t]
\centering
\caption{An overview of different Uncertainty Measures that capture predictive uncertainty/confidence from a distribution over probabilities.\\}\vspace{-1mm}
{\footnotesize
\adjustbox{max width=\linewidth}{%
 \begin{tabular}{l l l l l}
        \toprule
        Name & Formula & Intuition & Ale UQ & Epi UQ \\
        \midrule

        Class Probability \cite{borovac_calibration_2022}  & $\mathbb{P}(p) = \bar{p}_{\bar{c}}$ & Mean probability of predicted class & \checkmark & \checkmark \\

        Predictive Entropy \cite{milanes-hermosilla_monte_2021} & $\mathbb{H}_\text{pred}(p) = - \sum_c \bar{p_c} \log{\bar{p_c}}$ & Uncertainty in mean prediction & \checkmark & \checkmark \\

        Probability Variance \cite{zanna_bias_2022} & $\mathbb{V}_{\bar{c}}(p) = T^{-1} \sum_t (p_{\bar{c}t} - \bar{p}_{\bar{c}})^2 $ & Variance of the predicted probability & & \checkmark \\

        Expected Entropy \cite{xia2023benchmarking, smith2018understanding}  & $\mathbb{H_E}(p) = -T^{-1}\sum_t\sum_c p_{ct} \log{p_{ct}}$ & Average uncertainty for each prediction & \checkmark &  \\

        Mutual Information \cite{milanes-hermosilla_monte_2021} & $\mathbb{I}(p) \approx \mathbb{H}_\text{pred}(p) - \mathbb{H}_E(p)$ & Information gain from new sample & & \checkmark \\

        Margin of Confidence \cite{milanes-hermosilla_monte_2021} & $\mathbb{M}(p) = T^{-1}\sum_t p_{\bar{c}t} - \max_{\substack{c' \neq c}} p_{c't}$ & Average distance to second class & \checkmark & ?

        \\
        \bottomrule
\end{tabular}}
\begin{flushleft}
We consider some number of forward passes $t \in T$. 
We denote some number of classes $c \in C$. 
A given probability for a class $c$ on pass $t$ is then $p_{ct}$. 
The average probability of a class $c$ over all passes $T$ is denoted $\bar{p_c}$. 
To denote the highest probability class after averaging over $T$ we use $\bar{c}$. 
Lastly, $f_{\bar{c}}$ is the number of passes in $T$ where 
$p_{\bar{c}t} = \max_c p_{ct}$.
\\

\end{flushleft}}
\label{tab:uq_measures_overview}
\end{table*}

    \subsection{Information Theoretic measures of uncertainty}\label{sec:disentangling-entropy}

    \citet{smith2018understanding} describe a set of three complimentary measure of uncertainty rooted in Information Theory.
     
    We first consider Predictive Entropy, which measures the total amount of uncertainty over the probabilities of all classes. This is also a method commonly used for single-point Neural Networks. It is functionally equivalent to class probability for a binary classification task, but for more classes it also considers the amount of uncertainty remaining in the other classes. 
    
    Predictive Entropy\footnote{While the current work strictly defined this as predictive entropy, some works refer to this simply as entropy. Expected Entropy will sometimes also simply be referred to as entropy or Shannon entropy. In this work we consistently keep these distinct.} is given as:
    \begin{equation}
        \mathbb{H}_{\text{pred}}(p) = - \sum_c \bar{p}_c \log \bar{p}_c
    \end{equation}
    Variations of this include normalizing the entropy by dividing it by $\log(C)$ or taking $1 - \mathbb{H}_{\text{pred}}$ to get a confidence measure instead of an uncertainty measure \cite{phan_sleeptransformer_2022}.

    By capturing the total uncertainty, Predictive Entropy responds to both aleatoric and epistemic uncertainty. I.e. it is high when aleatoric uncertainty is high, or when epistemic uncertainty is high. Two more Information Theoretic measures have been proposed to disentangle these two sources of uncertainty.

    The Mutual Information between a model's parameters $\omega$ and a new labelled sample $\{x, y\}$ gives the amount of information gained by knowing the label of that sample, relative to what was already known by the model's parameters. Since this may be considered equivalent to epistemic uncertainty \cite{smith2018understanding} we get an intractable epistemic uncertainty measure:
    \begin{equation}
        I(\omega, y | D, x) = H[p(y|x, D)] - \mathbb{E}_{p(\omega|D)} H[p(y|x, \omega)]
    \end{equation}
    This can be approximated by sampling from the posterior distribution:
    \begin{equation}
        \mathbb{I}(p) \approx \mathbb{H}_{\text{pred}}(p) + T^{-1}\sum_t\sum_c p_{ct} \log{p_{ct}}
    \end{equation}
    These terms can be reordered as shown by \citet{mukhoti2021deep} into:
    \begin{equation}
        \underbrace{\mathbb{H}_\text{pred}(p)\vphantom{\sum_t}}_{\text{total}} \approx \underbrace{\mathbb{I}(p)\vphantom{\sum_t}}_{\text{epistemic}} - \underbrace{T^{-1}\sum_t\sum_c p_{\log{p_{ct}}}}_{\text{aleatoric if ID}}
    \end{equation}
    The latter part is referred to as \textit{Expected Entropy}, which is a measure of aleatoric uncertainty.

    This disentangling of Predictive Entropy into Mutual Information and Expected Entropy is well established in Computer Vision literature \cite{wimmer2023quantifying}, but we found surprisingly little traction for biosignal applications. Only \citet{zhang2024cardiac} used this set of complementary measures, though using only Predictive Entropy is more common \cite{aseeri_uncertainty-aware_2021, jahmunah_uncertainty_2023, phan_sleeptransformer_2022, milanes-hermosilla_robust_2023, xia2023benchmarking, deng_eeg-based_2023, fawden2023uncertainty, rahman2023quantifying}. 

    Note that this method of disentangling has been challenged \citet{wimmer2023quantifying, mucsanyi2024benchmarking, de2024disentangled}, showing that they do not result in a clean separation of aleatoric and epistemic uncertainty.

    \subsection{Margin of Confidence}
    Lastly, \citet{milanes-hermosilla_monte_2021} proposes the Margin of Confidence as an intuitive uncertainty measurement. This ad-hoc measure looks at the average distance between the probability of the predicted class and the class with the next highest probability. Note that while the predicted class is taken over the average from the forward passes $\bar{c} = \argmax_{c\in C} \bar{p}_c$, the second-highest is chosen on each sample. This means that in some forward passes, the second-highest probability $\max_{\substack{c' \neq c}} p_{c't}$ is actually higher than the probability of the predicted class $p_{\bar{c}t}$. 

    In its full form the Margin of Confidence is given as:
    \begin{equation}
            \mathbb{M}(p) = T^{-1}\sum_t p_{\bar{c}t} - \max_{\substack{c' \neq c}} p_{c't}
    \end{equation}
    \citet{milanes-hermosilla_monte_2021} used the Margin of Confidence to separate correctly and incorrectly classified predictions. They found that the Margin of Confidence had a greater Bhattacharyya distance between the correctly and incorrectly classified predictions than Mutual Information, Predictive Entropy and Probability Variance, but replications with other models, UQ methods and data are needed.

    \subsection{Recommendations for Uncertainty Measures}

    We find that the uncertainty measures used in the biosignal literature are often ad-hoc, lack thorough argumentation and are sometimes underspecified. We argue that future work should always specify how they measure uncertainty to ensure reproducibility.

    All uncertainty measures described here can be applied to any Uncertainty Quantification method that yield distributions over class probabilities (all Bayesian Neural Networks, Evidential Deep Learning, Early Exit Ensembles, VAEs, ADF, DUL and SDE-Net). It is not known whether certain measures are better for certain models. Measures of uncertainty are not sufficiently studied, and most papers use one without further justification or comparison. Only \citet{milanes-hermosilla_monte_2021} compares different measures on their ability to separate correct from incorrect classifications, finding that in their case Margin of Confidence performs best, but this cannot be generalized from. 

    Overall, it is recommend to use a measure that estimates aleatoric uncertainty, epistemic uncertainty, or both, depending on the task. In Computer Vision the established method of uncertainty measures for classification is using Predictive Entropy, Expected Entropy and Mutual Information. While this has substantial limitations \cite{wimmer2023quantifying, de2024disentangled, mucsanyi2024benchmarking}, we find that it is currently the best established approach. This gives a measure of total uncertainty, epistemic uncertainty and aleatoric uncertainty, though we caution that this disentanglement cannot be fully trusted. Instead, they should be used as \textit{best estimates}, rather than true predictions. For regression the aleatoric variance or the epistemic variance would be a best estimate \cite{jin2023uncertainty}. 

    Additionally, we believe that the class probability and the class variance are easy to interpret by clinicians, and are therefore most suitable when uncertainty estimates are used in a decision support system.

\section{Uncertainty Use Cases}\label{sec:uq-usecases}

    When applying Uncertainty Quantification to a biosignal application there should always be some purpose to the uncertainty estimation. Different ways of using uncertainty put different expectations on it, and the way uncertainty is used in biosignals comes with some special considerations.

    It also comes with different ideals for which kind of uncertainty should be used. We provide an overview of which (theoretical) uncertainty measure is most fitting for which use case in Table \ref{tab:usecase_per_uncertainty}. While there are no guarantees that measures for epistemic uncertainty only predict epistemic uncertainty, a paper should at least include the appropriate uncertainty for the appropriate task. We found that this is not always well understood in the biosignal literature, so this overview may help authors and reviewers.

    \begin{table}[t]
        \centering
        \caption{Various Uncertainty Use Cases grouped in their required type of uncertainty. In general, methods that need either aleatoric or epistemic uncertainty may still do well with a mixture of both. Rejection is split into rejection when the data is in-distribution (ID), or out-of-distribution (OOD), relative to the training data.}\vspace{1mm}
        \adjustbox{max width=\linewidth}{%
        \begin{tabular}{lll}
            \toprule
            Aleatoric & Epistemic & Both \\
            \midrule
            Feature & Active Learning  & Interpretability  \\
            Rejection (ID) & Model Pruning  & Social Bias   \\
             & Data Augmentation & Soft Voting  \\
             & Rejection (OOD) &  \\
             \bottomrule
        \end{tabular}}
        \label{tab:usecase_per_uncertainty}
    \end{table}

    \subsection{Rejection Methods}
        The most common use for estimating uncertainty is to be able to not make a prediction when the likelihood of that prediction being wrong is too high. 34\% of papers in this review use a measured uncertainty to reject difficult samples from the testing data. 
        
        Below we highlight how this impacts evaluation, the choice of uncertainty measures, and implementation in a biosignal context.

    \subsubsection{Evaluating Rejection methods}\label{sec:uncertainty_as_classification}
        A common technique used to evaluate uncertainty quantification for rejection methods is setting a threshold against uncertainty and observing an increase in accuracy and a decrease in coverage \cite{mohamed_detection_2005, fiorillo_deepsleepnet-lite_2021, harper_bayesian_2022, phan_sleeptransformer_2022, lin_robust_2023, lin_reliability_2022}. This framework considers uncertainty as a tool to improve classification performance, instead of having uncertainty as an inherent goal. While some works set a single threshold against uncertainty \cite{mohamed_detection_2005, fiorillo_deepsleepnet-lite_2021, lin_robust_2023, } we recommend a range of thresholds \cite{harper_bayesian_2022, phan_sleeptransformer_2022, lin_reliability_2022}, as the right balance between coverage and accuracy is typically not well established and a comparative analysis after a single threshold is not possible. 

\begin{figure}[t]
            \centering
            \includegraphics[width=\columnwidth]{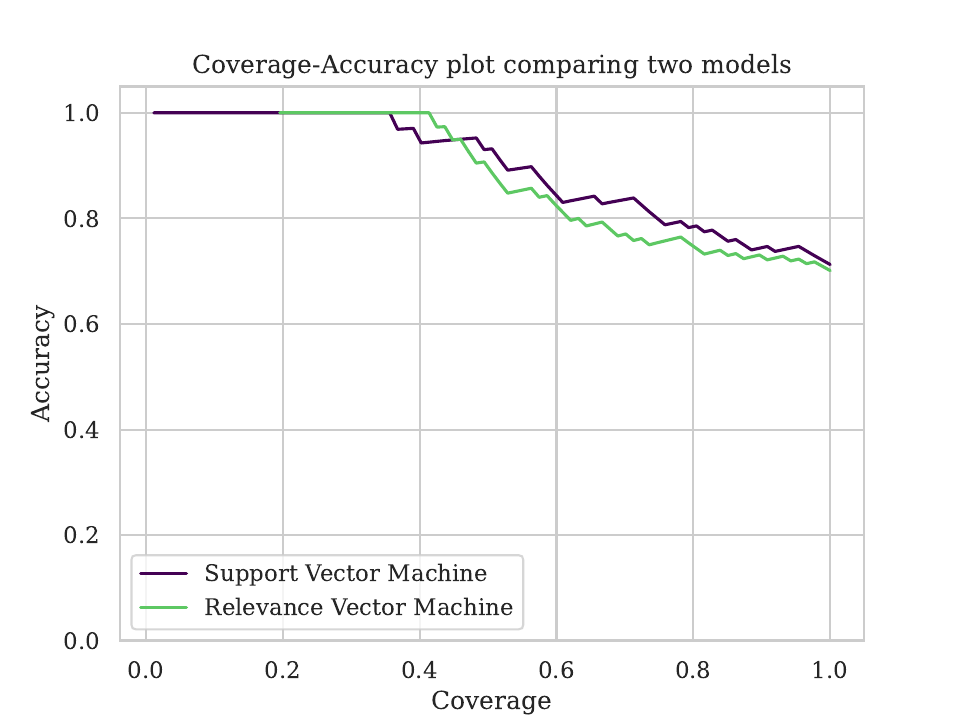}
            \caption{Example plot showing the trade-off between coverage and accuracy for two EEG Motor Imagery classifiers. The plot shows that both models have very similar accuracy without rejection (coverage at 1.0), but that the Support Vector Machine has a better accuracy-coverage trade-off.}
            \label{fig:coverage-accuracy}
        \end{figure}

        Instead, coverage-accuracy plots as visualised in Figure \ref{fig:coverage-accuracy} may be used to assess the reject-performance of a model. By going over all possible thresholds, this plot shows the options for balancing coverage and accuracy, which may be used for comparing models.

        The alternative framework is to consider uncertainty as a classification task, where the goal is to classify whether a prediction will be correct or incorrect \cite{milanes-hermosilla_monte_2021, aseeri_uncertainty-aware_2021, lin_reliability_2022, jahmunah_uncertainty_2023, barandas_evaluation_2024, milanes-hermosilla_robust_2023}. This results in the common classification metrics area under the ROC-curve \cite{huang2019evaluating}, but specifically for rejection.

        We recommend using the accuracy/coverage curve to explain the possible behaviour a classifier with rejection can exhibit, whereas a separate task-ROC and a rejection-ROC may give better insight into the individual components. 

        The results from \citet{jahmunah_uncertainty_2023} indicate a limitation of rejection with standard classification metrics. Their results show that even with large noise for an ECG classification task, the ECG is sometimes guessed correctly even when the model should be uncertain. Therefore, considering uncertainty as a classification task will inflate the number of false negatives and thus underestimate the rejection performance.

    \subsubsection{Choice of Uncertainty Measure}
         Both aleatoric and epistemic uncertainty can contribute to a risk of predictions being wrong, so total uncertainty would theoretically be optimal. However, in practice it may be that a measure of only aleatoric or epistemic uncertainty might work better. We recommend considering measures of aleatoric, epistemic and total uncertainty and seeing which performs best.

        Most works pick one uncertainty measure and do not actively compare them. Only \citet{fiorillo_deepsleepnet-lite_2021} made such a comparison, by considering both average class probability and probability variance as the uncertainty to use for rejection. They found the accuracy improved most under rejection with average class probability, across multiple datasets. However, it may be possible that other measures work better when there is more epistemic uncertainty involved.  %

    \subsubsection{The Rejected Samples}
    In rejection methods it is worth contemplating what happens to the samples that are rejected. \citet{van_gorp_certainty_2022} suggests that under epistemic uncertainty a clinician could re-assess the data, while under aleatoric uncertainty a re-recording of the electrodes would be needed instead \cite{belen_uncertainty_2020}, or even alternative tests \cite{larsen2023new}. However, we caution that current predictions of aleatoric and epistemic uncertainty are not sufficiently separable to implement such systems \cite{de2024disentangled}.
    
    Implementations where predictions are made and used in real-time require a well-considered behaviour for rejected cases. Machine Learning with rejection should consider how the rejected samples impact the larger clinical diagnosis system and what the outcome will be for patients who are rejected by the classifier.

    \subsection{Uncertainty for Interpretability}
        Uncertainty is sometimes proposed as a method to alleviate the black-box problem of Neural Networks~\cite{phan_sleeptransformer_2022}. %
        By presenting uncertainty, a model is able to show that a given prediction may not be correct, which can make the clinician more confident in trusting the certain predictions from a Machine Learning system. 

        Determining what good communication of a quantified uncertainty is can be difficult. %
        
        Research on scientific visualization of uncertainty is available \cite{bonneau2014overview, potter2012quantification}, but is not interweaved with the reviewed literature and does not demonstrate how to present different measures of uncertainty. Specifically clinical interpretation of uncertainty is critical, as it may affect the quality of a diagnosis or the adoptability of Machine Learning methods. For some ECG applications time-sensitivity is given as a factor affecting manual diagnosis \cite{jahmunah_uncertainty_2023}, so the interpretation of an uncertain prediction may be subject to time constraints in such cases.

        Conformal Prediction may be a good approach for communicating uncertainty to a user, by specifically not quantifying an amount of uncertainty. Giving a set of multiple possible diagnoses has been shown to improve diagnosis accuracy compared to giving a most-likely diagnosis paired with a probability \citet{folgado2025conformal}. 

        In standard classification tasks an accepted way of presenting a quantified uncertainty is by reporting an accurate class probability. A predicted class probability that accurately corresponds to the true probability of a class (even under epistemic uncertainty) can be mathematically interpreted and gives a well-defined and well-understood measure of uncertainty. Expected Calibration Error (ECE) has been used to capture this goal in a metric \cite{borovac_calibration_2022, xia2023benchmarking, campbell_robust_2022}. 

        \begin{figure}[t]
            \centering
            \includegraphics[width=\columnwidth]{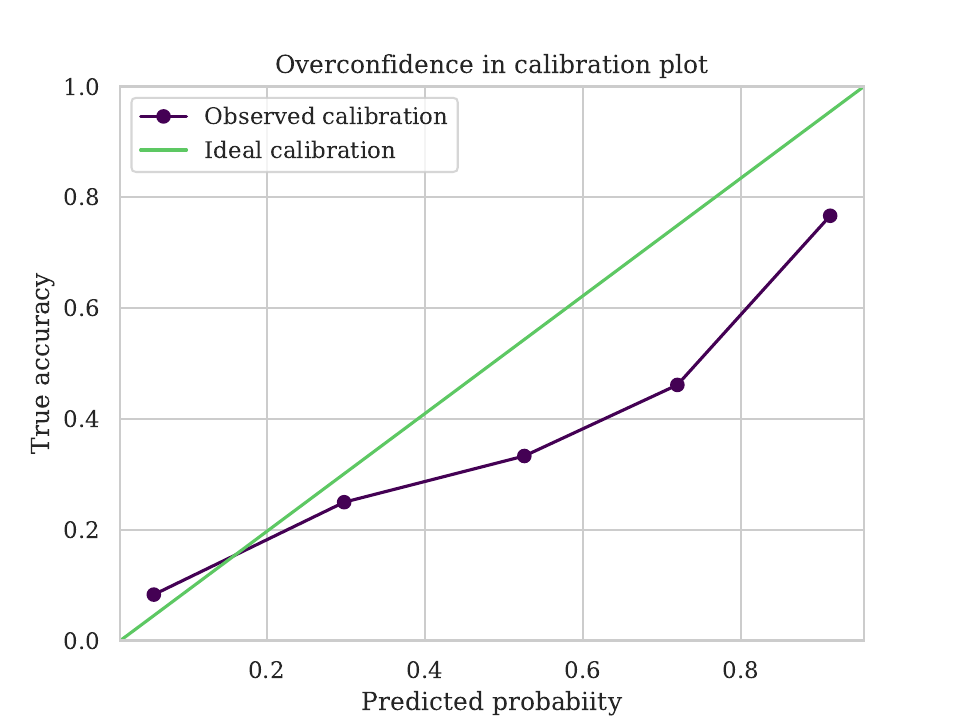}
            \caption{Example calibration plot for an EEG Motor Imagery classifier. The predicted probability is consistently higher than the true probability of being correct. This means the model is overconfident. Note that a plot like this is only reliable with sufficient samples with different predicted probabilities. }
            \label{fig:calibration_plot}
        \end{figure}

        \begin{figure*}[!htb]
        \centering
        \subfigure[Explicit prediction and uncertainty]{
          \begin{minipage}[t]{0.31\linewidth}
          \centering
          \includegraphics[width=\linewidth]{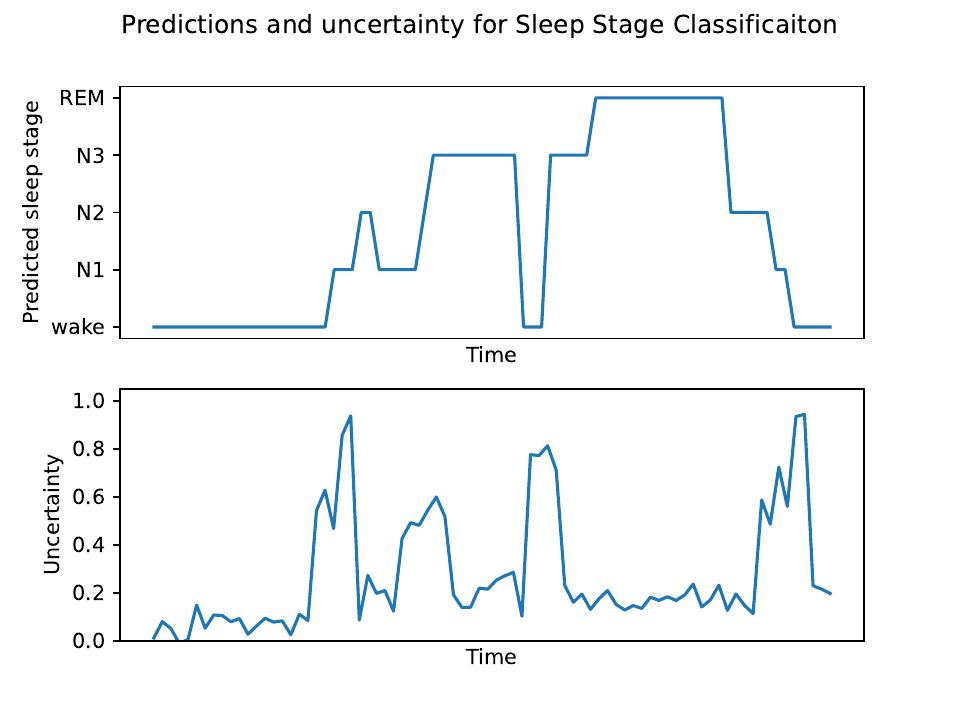}
          \end{minipage}
        }
        \subfigure[Explicit uncertainty and implicit prediction]{
          \begin{minipage}[t]{0.31\linewidth}
          \centering
          \includegraphics[width=\linewidth]{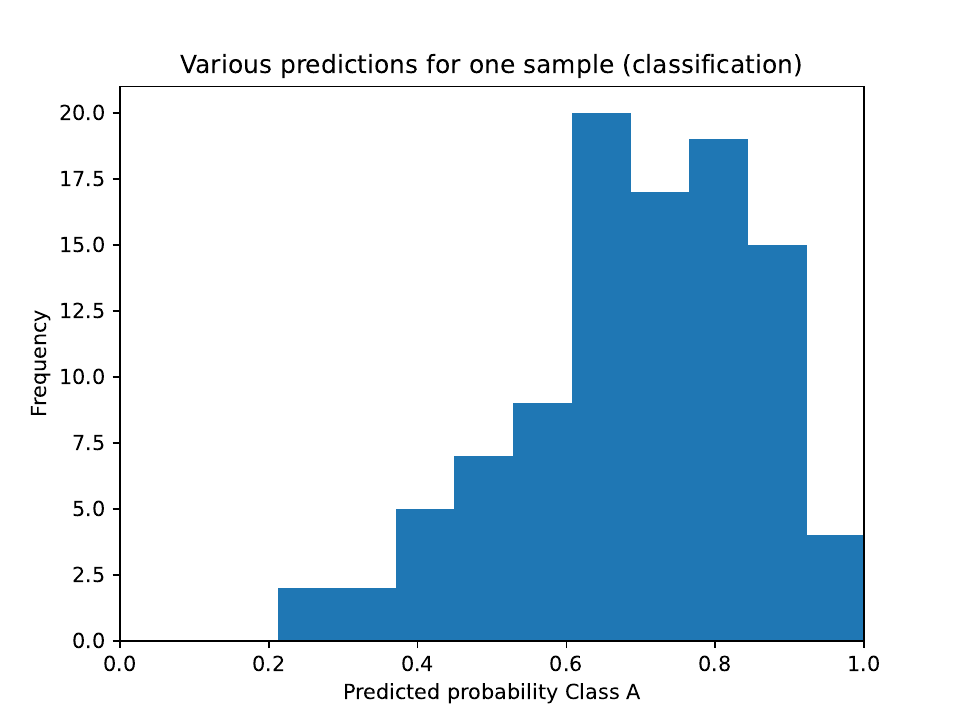}
          \end{minipage}
        }
        \subfigure[Implicit prediction and uncertainty]{
          \begin{minipage}[t]{0.31\linewidth}
          \centering
          \includegraphics[width=\linewidth]{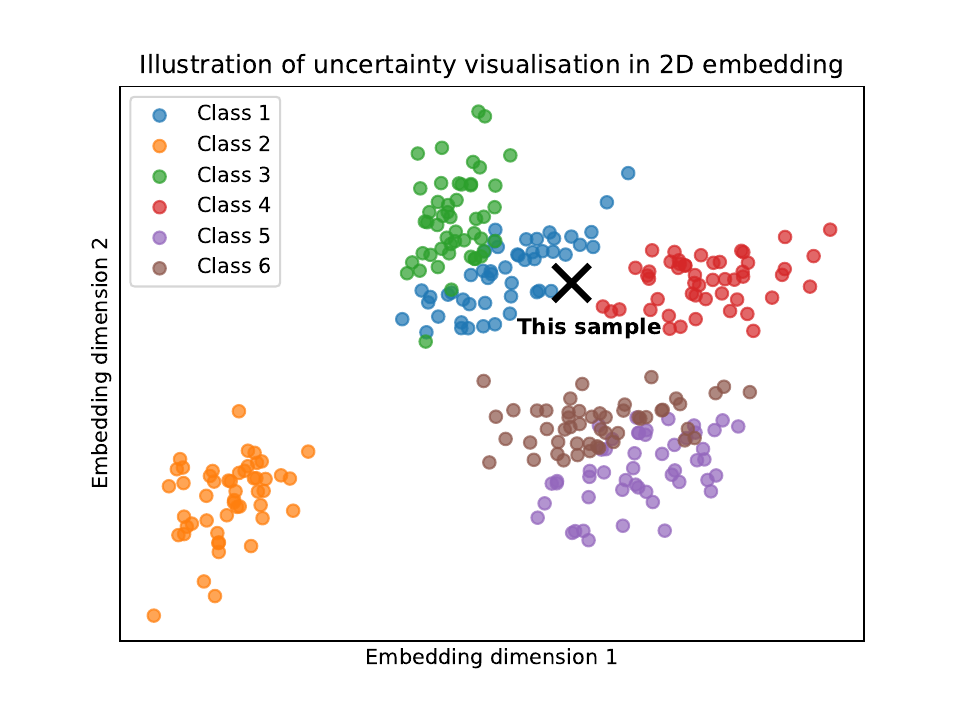}
          \end{minipage}
        }
        \caption{Three different general approaches to visualising uncertainty. The first plot specifically shows a separate prediction and a degree of uncertainty. The second plot explicitly shows the uncertainty, but leaves the predicted probability to be determined by the user. The third plot shows an embedding of the sample, but does not explicitly give a prediction nor an uncertainty.  }
        \label{fig:uncertainty_visualisations}
        \end{figure*}

        ECE is a common method for evaluating uncertainty quantification. It measures the difference between a predicted probability for a classification and the actual observed probability on a validation set \cite{niculescu2005predicting}. This is often visualised with a calibration plot as shown in Figure \ref{fig:calibration_plot}. While Expected Calibration Error measures directly the correspondence between predicted probability and true probability, it does not show what is the cause. A consistently over-confident or under-confident classifier can both have a bad ECE. We recommend additionally looking at Net Calibration error \cite{groot2024overconfidence}, which measures the over/under confidence in isolation.

        ECE is only defined for classification problems. For regression problems a similar method exists called ENCE \cite{levi2022evaluating}. In this method similar bins are made, but instead of comparing accuracy with predicted probability, it compares root mean-squared error to the predicted root mean variance. This can also be evaluated with plots similar to the calibration plot in Figure \ref{fig:calibration_plot}. 
        
        Currently, we recommend ECE as a metric to evaluate predicted probabilities when those probabilities are to be interpretable to a clinician, but research on how clinicians interact with predicted uncertainty is lacking. It may be that uncertainties are easier to interpret if presented as natural language as in \citet{mendoza2023deep}, which would result in different evaluation criteria. Additionally, it may be necessary to evaluate uncertainty on data that is recorded with the same hardware, in the same clinic and by the same people as where it would be implemented, as this may introduce more epistemic uncertainty.

        \subsubsection{Visualizations of Uncertainty}
            Interpretation of uncertainty may be further improved by having an appropriate visualization that aligns with the specific biosignal task.

            We generalised the reviewed approaches into three different categories that are shown in Figure \ref{fig:uncertainty_visualisations}. We found visualisations that makes a distinction between the prediction and the (epistemic) uncertainty, visualisations that show all possible predictions (e.g. as a histogram) to show uncertainty while leaving the prediction implicit, and visualisations that offer insight into a sample without explicitly quantifying the prediction or the uncertainty. The reviewed visualisations do not have theoretical or empirical arguments for their design, but by defining the framework we offer some grasp on otherwise varying visualisations. 
            
            The specific design details vary depending on the exact task and application context. We discuss those details and variations found in the reviewed literature below.

            \citet{gill_multicenter_2021} uses a CNN with MC-Dropout to classify lesional voxels in patients with focal cortical dysplasia. The results are presented by a map of class probability voxels (predictive uncertainty) and a separate map of probability variance voxels (epistemic uncertainty). This gives an explicit separation between the prediction and the epistemic uncertainty. It is then up to the user to combine these two sources of information. 

            \citet{bekhti_hierarchical_2018} proposes a Markov Chain Monte Carlo (MCMC) approach to solve the inverse problem. The MCMC sampling results in multiple sparse solutions, where the agreement between solutions is interpreted as uncertainty. By presenting a heatmap of the source localization solutions on 3D brain renderings they allow the reader to interpret the level of uncertainty based on the relative density and the total spread of solutions. This also allows readers to involve their prior knowledge about neuroanatomy implicitly by contrasting the certainty of the predictions against prior knowledge. 

            \citet{phan_sleeptransformer_2022} shows a method to support EEG-based sleep classification. They show a timeseries of the predictive entropy, the stacked class probabilities and the classifications above each other. To improve readability they highlight the parts where confidence drops below a given threshold. This is used to show how uncertainty is highest during stage transitions. Representing uncertainty over the timeseries, in combination with the original signal will let a Machine Learning system work as an effective decision support system for biosignal analysis. 

            A more generalisable method is given by \citet{costabal_machine_2019}, who present a histogram of the whole distribution of class probabilities. This allows readers to intuitively asses central tendencies, spread and skew, and generalises naturally to visualisations for regression.

            An exceptionally interesting approach to dealing with the interpretation of uncertainty is suggested by \citet{van_de_leur_interpretable_2021}, where a VAE embedding of an ECG is reduced to 2 dimensions using Principal Component Analysis. A cardiologist is presented with the embeddings of known diagnoses. This allows them to determine a measure of uncertainty based on a more fluid notion of vacuity, dissonance, aleatoric or epistemic uncertainty. By not trying to quantify uncertainty, but instead allowing the cardiologist to assess uncertainty, this method aims to make a diagnosis more interpretable. 

    \subsection{Uncertainty as an Instrumental Goal}
        All other usecases of uncertainty we found were using uncertainty to improve some other task, such as Active Learning, and pruning in a more complicated classifier pipeline. For these cases, evaluating the uncertainty specifically has limited relevance as it is not an output from the system. Instead, the impact of uncertainty should be measured based on how it helps with the downstream task. For these aspects, the specific relation to biosignals is somewhat limited, as these may also be applicable to other tasks. However, we highlight these as this topic gets little attention in literature reviews focused on methodology. 

        We outline the setups that use uncertainty to improve some other outcome to demonstrate possible setups, and to illustrate the usefulness of uncertainty. 

    \subsubsection{Uncertainty as a Feature}
         The most direct use of uncertainty is as a feature for subsequent Machine Learning tasks. For example, ~\citet{stoean_automated_2020} attempts to detect presymptomatic spinocerebellar ataxia using electrooculography. They observe the saccadic eye movements in healthy, sick, and presymptomatic participants. Healthy participants show a sudden eye movement with nearly instant acceleration and deceleration. Sick participants can show more chaotic movement with slower acceleration and speed. Presymptomatic participants can show a decrease in control, speed and rate of acceleration. Since there is a lot of variation between participants and each saccade, 85 saccades are recorded for each participant, and classified with an ensemble of Deep Neural Networks using MC-Dropout. The 3 class probabilities and the 3 class standard deviations for all 85 saccades were used for a decision tree classifier. The system was able to classify sick and healthy participants quite well, and performed acceptably at classifying presymptomatic participants.

        When uncertainty is used as a feature for another Machine Learning model the constraints of what a good uncertainty is are loosened. Uncertainty may be expressed with multiple uncertainty measures, and over or underconfidence will not have an impact on the system.

    \subsubsection{Uncertainty to Control Social Bias}
        As fairness and negative social biases are a growing concern in Machine Learning, \citet{zanna_bias_2022} present a rather unique usecase for uncertainty quantification. They propose a Multi-Task Learning method using Uncertainty Quantification to reduce social bias while classifying periods of anxiety from ECG features. The bias mitigation strategy uses a separate output that attempts to classify whether the samples belong to a person from an unprivileged demographic group. 

        The model is trained for 100 epochs, with the weights being saved every 5 epochs. After training, the model with the highest average epistemic uncertainty (probability variance) on the demographic-classification and the lowest average uncertainty on the anxiety-classification is selected. The model performing poorly at demographic classification should not have features in the latent representation to capture demographic classification. The authors showed that this minimized bias, but this did come at a loss in model performance.

        While this method is still somewhat ad-hoc, it paves the way for future methods in minimizing social bias through uncertainty quantification. Future research may focus on forms of adversarial training, so that an anxiety model will try to optimize the anxiety classification while under an ongoing constraint of having no features that may be used to infer the demographic class. The different effects of aleatoric and epistemic uncertainty are also worth exploring here.

    \subsubsection{Bayesian Active Learning}
        Under Active Learning training samples are iteratively selected by the epistemic uncertainty that the model has about that sample \cite{gal2017deep}. These methods are proposed for situations where insufficient labelled training data is available, and manual labelling of data is expensive. Active Learning starts with a model trained on very little data, and observes the uncertainty it has on the unlabelled data. The most uncertain samples are then manually labelled by an \textit{Oracle}: a system that produces the ground truth labels. This Oracle can be the expert annotations, but may also be additional (expensive) testing to establish a better ground truth. We found three different ways this is used for uncertainty.

        \citet{wabina_neural_2022} compared their Neural Differential Equation approach to a BNN trained with Active Learning. BNNs that use Active Learning can use their (epistemic) uncertainty to indicate about which samples they are uncertain. Remarkably, the best performance was actually observed by predictive entropy (total uncertainty rather than Mutual Information (epistemic uncertainty), presumably due to poor uncertainty disentanglement.

        \citet{vavaroutas2023uncertainty} instead uses Active Learning to guide their Data Augmentation process for ECG and EEG classification tasks. They add Data Augmentations to an existing dataset, theorise a scenario with unlabelled samples, and use Active Learning to achieve acceptable performance with only 20 annotated samples. We believe that this is a promising direction, but that care should be taken to ensure that if augmentations need to be annotated by clinicians, then those augmentations should be specifically designed to maintain the integrity of a realistic signal. Clinicians might not be able to annotate a horizontally flipped ECG. 

        Lastly, \citet{fawden2023uncertainty} uses a method similar to Active Learning to reduce the size of the dataset to train the model on. They show that this reduces the computational cost of transfer learning, which may be important for edge devices where the biosignal recording may be privacy sensitive and must be used to train a model locally. 

        Overall we find that Active Learning is a promising avenue, although more work is needed to understand the downstream impact of using Bayesian Active Learning.

    \subsubsection{Miscellaneous use cases for uncertainty}
        Two works propose novel ways to use uncertainty for Brain-Computer Interfaces. As part of their UNCER model, \citet{duan_uncer_2023} uses uncertainty to assess the quality of data augmentation. They consider data augmentation as a method to reduce uncertainty to unseen corruptions. 

        For a P300 speller \citet{ma_bayesian_2023} look at model uncertainty, not only in terms of how it affects predictive uncertainty, but also in what it says about the model. They argue that weights with a poor signal-to-noise ratio are redundant. With this method they were able to prune 75\% of the weights without decreasing the F1 score. In the single-point model any amount of pruning would result in a (slight) decrease in F1 score.

        Additionally, \citet{ma_bayesian_2023} used the predicted probability for a special soft-voting strategy. In P300 spellers each letter is flashed several times, and a classifier tries to identify a P300 wave. By using the probability of a P300 wave their Bayesian CNN outperformed an equivalent single-point model. This strategy of voting with probabilities, rather than with discretised predictions is similar to Soft Voting in Machine Learning ensembles.

    \subsection{Recommendation for use cases}
        We advise future work on applications of Uncertainty Quantification to specify what purpose of uncertainty estimation they are considering. One may consider a rejection scenario, a decision support system, or using uncertainty as an instrument to achieve some other goal. 

        If the goal of uncertainty is to achieve good rejection, appropriate evaluation should use the accuracy/coverage curve or consider rejection as a classification task and present an ROC-curve. Reporting results with a single threshold is not sufficient, as it cannot be interpreted. We also find that it is well established that rejection gives some benefit, so studies should focus on comparing methods to achieve the most benefit, or investigate how to deal with rejected samples in a clinical setting. 

        Works focusing on uncertainty to improve interpretability can evaluate their methods using ECE, NCE and the Brier score, though a rejection ROC-curve may be a good addition. Foundational research on how clinicians interpret predictive uncertainty, how to communicate it, and how to visualise it are also needed.

        When uncertainty is used as an intermediary, for example for pruning, Active Learning, or as a feature for another model, there are fewer constraints to the uncertainty quantification, and the evaluation does not need to be as thorough. Instead, evaluation should look at how uncertainty estimates affect the downstream task performance.

\section{Guidelines for Adding Uncertainty Quantification}\label{sec:how_to_build_uq}

The review covered various methods for obtaining quantified uncertainties and presented methods which people have been using uncertainty for. Based on these findings, we aim to conclude a guideline on how to implement uncertainty quantification for a Machine Learning task on biosignal data. There is no singular solution or decision tree that works best for all cases. Nonetheless, we provide an outline below of decisions to make for researchers using a Machine Learning system for a biosignal task that are interested in using Uncertainty Quantification. These instructions should be taken with a critical eye and may be subject to disagreement. Still, it provides a starting point from which further methodologies may be constructed. 

We start by considering the cost of adding Uncertainty Quantification to a Machine Learning task. After this the first step will cover the uncertainty quantification methods, which is mostly guided by your choice of Machine Learning model and computational constraints. Second is the choice of uncertainty measure, which is chosen on the constraints of the uncertainty usecase. The last step is the evaluation. Depending on the uncertainty usecase, different evaluation methods align best with the specific goal. Lastly, we discuss some sanity checks to validate that the uncertainty quantification works as intended. 

\subsection{Choice of Uncertainty Quantification Method}
Knowing when your model's predictions are likely to be wrong, and a hint of why they might be wrong, can be quite valuable. However, there is always a price to pay.

For MC-Dropout and Deep Ensembles this price is computational cost. MC-Dropout requires many forward passes, so the cost of inference might increase 100 times. Deep Ensembles require training several models, which means training cost may increase 5 times. At inference, this also requires having enough memory for 5 models.

However, these methods do not result in a decrease in model accuracy. MC-Dropout converges to roughly the same prediction that a single-point model would have made after 100 forward passes \cite{valdenegro2022deeper}, and ensembles are well established at improving model accuracy \cite{sagi2018ensemble}.

Methods that optimise a model for uncertainty (such as Variational Inference, Prior Networks, Evidential Machine Learning and Variational Autoencoders) are at risk of decreased model accuracy. Since the model is now optimised towards two tasks simultaneously, this may have a negative effect on the predictive performance. However, this is not guaranteed as multi-task learning leverages a similar mechanism to improve predictive performance \cite{zhang2021survey}. 

Post-hoc calibration does not directly have a substantial computational cost, nor does it directly affect the model predictions. However, doing post-hoc calibration requires data to do the calibration on, which generally cuts into the data available for training or testing. 

We generally recommend trying Deep Ensembles, MC-Dropout and a standard Neural Network and comparing their performances for the task at hand. If a five-fold increase in training cost or a 100-fold increase in inference cost a prohibitive only either Deep Ensembles or MC-Dropout may be a viable starting point. If well-calibrated uncertainties are a requirement, we recommend adding a post-hoc calibration method such as temperature scaling. 

When the computational cost is a large constraint, one might try Evidential Deep Learning or Early Exit Ensembles to further reduce computational cost.

If the base-model of choice is not a Neural Network there is little previous work available to build on. We recommend implementing Bayesian methods for standard Machine Learning models such as Bayesian Logistic Regression, Bayesian Linear Discriminant Analysis and Relevance Vector Machines as explained by \citet{princeCVMLI2012}, or doing bootstrap ensembling \cite{larsen2023new}. These methods have the ability to incorporate epistemic uncertainty, which is otherwise neglected. 

\subsection{Choice of Uncertainty Measure}

There is fairly limited literature on regression with biosignals \cite{costabal_machine_2019, zhang2023knee, zhang2023neuromusculoskeletal, wabina_neural_2022}, but we recommend from our experience two measures of uncertainty for regression: the variance of the prediction, or the 95\% Confidence Interval. Measures of variance may be well suited for rejection systems, as they present a scalar uncertainty that can be thresholded against. Confidence Intervals may be preferable for human interpretation as they give a notion of likely possible values.

For classification problems the current state-of-the-art is more conflicting. For rejection the predictive entropy, expected entropy, or mutual information may all be good options. While they theoretically correspond with total, aleatoric and epistemic uncertainty in practice this is not straightforward and we recommend trying all three.

Alternatively, the Gaussian Logits disentangling gives a predicted probability, aleatoric variance and epistemic variance, but this is less established in the current literature. Further research comparing these two methods of disentangling uncertainty 
for biosignals is needed.

For uncertainty to be interpreted by people (clinicians or users) a (well-calibrated) class probability is easiest to interpret. Epistemic uncertainty may be represented by the class variance, but would ideally be incorporated into a more uniform probability distribution.

When the purpose of the uncertainty is an intermediary multiple measures may be observed and combined with dimensionality reduction methods as needed. However, we expect that a combination of aleatoric, epistemic and mixed uncertainty measures will perform best. 

\subsection{Evaluating Uncertainty Quantification}
Whenever Uncertainty Quantification is considered as a tool to improve the outcome of a larger system, rather than as its own end-goal, the evaluation methods may need to be adjusted to the purpose for which uncertainty is used. Below take in each section a given uncertainty usecase, and discuss how to evaluate the uncertainty quantification for that usecase.

\subsubsection{Rejection}
If uncertainty is used in order to reject difficult samples,  the impact of uncertainty on the larger system may be directly measured with a coverage-accuracy plot as in \cite{phan_sleeptransformer_2022, lin_reliability_2022}. These systems all depend on setting a threshold, which is usually arbitrary. Therefore, it is better to create a plot that shows the outcome for all possible thresholds by plotting the coverage against the accuracy. Showing the coverage and accuracy only for a single threshold makes it hard to compare models when the distribution of the uncertainty measure shifts. 

However, these coverage-accuracy plots do not give direct insights into the Uncertainty Quantification performance per se. Gaining more insights into this may help improve the large system, rather than only evaluate it. For this, it may be worth casting the uncertainty as a classification task, so that regular classification metrics may be used. Be aware that this is typically an unbalanced task, where again the cost of false-positives and false-negatives is not well defined, so ROC curves may be a preferred approach. Since a perfect uncertainty measure is not able to provide perfect classification (as described in Section \ref{sec:uncertainty_as_classification}), it may be worth adjusting the metrics to give a more directly interpretable evaluation of the uncertainty. 

For both of these cases, it is worthwhile to use a good baseline to assess whether the Uncertainty Quantification method actually provides an improvement. Setting a threshold against a standard Neural Network with Softmax as uncertainty gives a fair baseline. 

\subsubsection{Interpretation}

While the rejection usecase does not demand a well-calibrated measure of uncertainty, this may be important for interpretation by a person. In this case the best approximation that can be given is that a predicted probability should align with the true probability. This can be measured by the Expected Calibration Error (or ENCE for regression \cite{levi2022evaluating}), which is therefore an acceptable metric for evaluating an uncertainty that needs to be directly interpreted. 

However, giving too many significant figures of a probability may give a false sense of precision, so it is possible that similar probabilities can be put in larger bins, which may even be mapped to natural language. In that case, the Expected Calibration Error is not ideal, as many small errors can have a substantial contribution to this metric, but may not actually affect the presented uncertainties. Instead, Maximum Calibration Error may be used, as this would ignore the small calibration errors and only focus on the large differences. 

For a thorough understanding of what works best for interpretability, human evaluation and user studies are needed. Both for the general problem of using uncertainty quantifying ML models, as well as for specific user groups and specific tasks. For supporting interpretability in medical decision making user studies should focus on the specific medical discipline of the user. 

\subsubsection{Intermediary Features}
When uncertainty is used as an intermediate, for example as a feature for a different model, or as an acquisition function for  Active Learning, it can be hard to identify which properties are required for an optimal uncertainty measure. 

ECE / ENCE may be used as a proxy for the quality of the uncertainty, but this is not specific to the usecase. Instead, the uncertainty method should be evaluated on the impact it has on the performance of the larger system. 

For any case of using uncertainty, it may be good to perform some sanity checks to ensure the uncertainty is behaving as intended \cite{valdenegro2021exploring}. For systems that are expected to measure epistemic uncertainty, one may try to create out-of-distribution data, and validate whether the epistemic uncertainty increases. To observe the quality of aleatoric uncertainty, one may look at the samples in the training data that are classified with high aleatoric uncertainty, to assess whether they align with the intuitions for aleatoric uncertainty. Alternatively, aleatoric uncertainty may be evaluated with relevant and realistic induced noise in the training data.

\section{Open Challenges}\label{sec:open-challenges}

We close the review by highlighting several open challenges of using uncertainty quantification for biosignals that warrant attention. Overall, while uncertainty quantification has been gaining traction, there are still multiple obstacles for adoption and under-explored areas. This paper removed some obstacles by providing an outline of how to add Uncertainty Quantification to a biosignal classifier in Section \ref{sec:how_to_build_uq}. We invite more researchers to incorporate Uncertainty Quantification methods into their models and the address remaining open questions, as discussed in this section.

\subsection{Interpretability of Uncertainty}
This review found 14 papers where the quantified uncertainty was explicitly or implicitly intended to be interpreted by a person, but none of them connected the uncertainty to thorough studies of how different representations affect uncertainty. \citet{gill_multicenter_2021} - for example - makes a visualization distinguishing predictive and epistemic uncertainty in FCD lesions detection, but it is not known how well such a visualization helps a clinician with identifying the true lesions and the false positives. \citet{mendoza2023deep} bins uncertainty estimates into natural language (including "Cannot rule out", "Consider" and "Possible") to be more intuitive, but the impact this has on interpretation is not yet known, and it may require different metrics for evaluating uncertainty.  

Previous research about how well clinicians can interpret probabilistic tests exists \cite{kostopoulou2022using, palfi2022algorithm}, but that is currently not tied to the way Uncertainty Quantification research is conducted. Research on what makes a well-interpretable (disentangled) uncertainty is needed, with an emphasis on designing visualisations. 

\subsection{Small Uncertainty Models for Biosignals}
Bayesian Neural Networks cover the majority of uncertainty quantification methods encountered in this review. These methods have been popularized in Computer Vision, where Deep Neural Networks are dominating the state-of-the-art. 

While Deep Learning has been gaining popularity and generating good results on large datasets \cite{somani2021deep}, its infamy for requiring large amounts of training data means many Biosignal models prefer shallower Machine Learning systems such as Support Vector Machines \cite{kawashima2017prediction} and Linear Discriminant Analysis \cite{yeh2009cardiac}. This review did not find much uncertainty quantification for such models, although they do exist (see \citet{princeCVMLI2012}). More research implementing uncertainty quantification on shallow models is needed, preferably with the ability to disentangle aleatoric and epistemic uncertainty, but minimally with the ability to capture a mixture of aleatoric and epistemic uncertainty. \citet{larsen2023new} provides a starting point with pseudo-bootstrap ensembles, but a more thorough analysis of uncertainty for such a model is needed.

\subsection{Appropriate Benchmarks for Uncertainty}
\citet{xia2023benchmarking} offers some benchmark data. They do this by introducing noise to existing biosignal datasets with the intention that uncertainty should go up as dataset shift makes the accuracy go down. They specifically model types of noise that could arise in practice, such as clipping, background noise (simulated with TV news audio), and missing segments. This gives a good set of uncertainties that may serve as a benchmark. However, this is still limited to shifts that are simple to simulate.

Epistemic uncertainty presents most realistically in cross-subject generalisability, rare comorbidities, or unusual erroneous recordings. By tailoring a dataset with these sources of epistemic uncertainty, we can improve the construct validity of UQ research. For designing such datasets we encourage looking at out-of-distribution detection datasets in Computer Vision as a starting point \cite{mukhoti2021deep}, but with clear attention to what is realistic in biosignals.

\subsection{Vacuity-Dissonance and Aleatoric-Epistemic}
Two frameworks for understanding uncertainty were encountered. The most common is the distinction between aleatoric (data) and epistemic (knowledge) uncertainty. However, the vacuity (absence of class features) and dissonance (contradicting class features) distinction could provide a more directly interpretable disentangling of uncertainty. It is not clear how these frameworks interact, and clarifying this may provide a more complete understanding of the uncertainty a model encounters. 

Future research may explore their interactions, their differences, and other interpretations of uncertainty that may be useful for biosignal classification tasks.

\subsection{Uncertainty in Regression}
Most of the reviewed literature focused on classification tasks, with only a few papers focused on uncertainty in regression. Methods for predicting, evaluating and communicating uncertainty in regression do exist, but since they are less prevalent, less is known about possible unique properties. There have been several extensive comparisons of UQ methods specifically using biosignals, but only for classification problems. 

Thorough comparison of regression methods with uncertainty, as well as a critical look at how these methods are evaluated, is still needed. As discussed, biosignals can suffer from high dimensionality, noise, and low sample size, which may have a specific impact on the quality of different regression methods and how they can be evaluated. 

Additionally, regression problems may come with unique challenges for communicating uncertainty in a medical setting. Real-time monitoring of vitals typically does not include a representation of heteroscedastic uncertainty. Research is needed on whether quantiles, variance, or histograms would make for usable and interpretable methods for uncertainty estimation in regression. The only work on interpretable uncertainty in regression we found was \citet{martinez_strategic_2020}, which looks at generating interpretable ECG with uncertainty estimates based on bio-impedance. However, they do not evaluate their method with users. 

\subsection{The Needs of Clinicians}
\citet{elul_meeting_2021} discusses the needs of clinicians in three concepts: estimating uncertainty, handling unknown classes, and detecting a failure to generalise.

Under the aleatoric-epistemic uncertainty framework, the \textit{estimating uncertainty} corresponds to aleatoric uncertainty, while both out-of-distribution unknown and known classes fall under epistemic uncertainty. In order to better address the clinical concerns, each of these problems may be addressed separately. While the path towards this is not known, the unification of aleatoric-epistemic and vacuity-dissonance uncertainties may provide a starting point.

\subsection{Using Uncertainty for Biosignal Applications}
56.6\% of the reviewed papers use uncertainty either for presenting a confidence with a prediction, or for rejecting difficult samples. However, there is an unknown number of other possible things that uncertainty quantification may be used for that need exploring. 

A promising purpose is to use uncertainty in an online setting while recording a biosignal. An increase in uncertainty may correspond with artefacts in the data, making uncertainty an artefact detector with possibly better properties than normal artefact classifiers. One advantage is that it may only detect artefacts that are obstructing a good classification, allowing it to tolerate artefacts in channels or at timepoints where they do not pose a problem for the specific task.

There may be many more unexplored opportunities to use estimated uncertainties when these uncertainty-enabled models are integrated in a task environment. Perhaps in a neurorehabilitation BCI the uncertainty may be used to support the patient in improving their movement attempts, or in situations where the labels may be erroneous an uncertainty measure is able to detect mislabeled training samples \cite{arriaga2023difficulty}.

\subsection{Informative Priors}
Variational Inference gives a modeler the option to specify a prior $p(\theta)$. This prior may be very helpful in training good Bayesian Neural Networks when data is limited. Efforts to cast domain knowledge into a probability distribution for $p(\theta)$ may be non-trivial, but this has the potential to improve these models. 

Alternatively, the prior $p(\theta)$ may also be learned on datasets similar to the task at hand \cite{shwartz2022pre}. 

\subsection{Rejected Samples}
We see that several works reject difficult samples to improve accuracy. In medical diagnosis systems the assumption is that these difficult samples may be offered to a diagnostician, so that their quality of diagnosis may not be compromised by mistakes in the Neural Network. However, it is unclear what the resulting diagnostic performance of the whole clinical system would be when combining the assessment of the doctor with the prediction of the Neural Network. It may be that they find the same samples difficult, 

In Breast Cancer and Tuberculosis screening some theoretical work with historical data has been done \cite{dvijotham2023enhancing}. Similar research may be done within the biosignal domain as a step towards implementing models with Uncertainty Quantification in the medical biosignal domain.

\subsection{Label Ambiguity}
Supervised Machine Learning considers the labels as \textit{ground truth}. However, in reality these ground truths may not be entirely accurate. This is often due to ambiguity or annotator error. To achieve appropriate estimates of uncertainty, the uncertainty of the ground truth should also be considered, but we found that this does not yet get enough attention. 

\citet{ju2022improving} demonstrates various methods for dealing with annotator disagreement on medical image classification. They demonstrate that the usual approach of establishing the ground-truth annotation by majority-vote is insufficient, and proposes a method that achieves better accuracy. 

Label ambiguity is especially common in biosignal analysis, as the ground truth often cannot be reliably established. \citet{zhang2024cardiac} found that models with highly confident (incorrect) predictions corresponded with error or ambiguity in ECG labels. They found a subject with two arrhythmia, but the original label only included one of them. In sleep stage classification there is often disagreement about the exact onset of a different stage \cite{phan_sleeptransformer_2022}, and class definitions might even change with differences in the gold standards \cite{danker2009interrater}. Generally it can be assumed that there is at least some label ambiguity in these biosignal datasets, but quantifying how much and for which samples is also important. Knowledge of label ambiguity can improve uncertainty estimation, and is valuable for further investigation.

\subsection{Foundation Models for Biosignals}
Language modelling and Computer Vision have gained substantial benefit from using foundation models. These are large models pre-trained on large dataset as a way to learn a general feature-extractor, that can be used for few-shot learning to achieve high performance on a task with very few specific training examples. The Large Brain Model (LaBram) \cite{jiang2024large} is such a large pre-trained model trained on 20 different EEG datasets. Instead of learning a specific task such as diagnosis or decoding, these large models are first trained on an arbitrary task, typically patch reconstruction, to ensure a comprehensive representation of the original signal \cite{gu2025foundation}.

Foundation models are an active and promising direction of research \cite{barmpas2025advancing, barmpas2025neurorvq, wang2025cbramod, tian2024foundation, mckeen2025ecg}, typically yielding improved task performance in small-data paradigms. However, their uncertainty quantification ability has not been studied in biosignal contexts. Methods such as Deep Ensembles and Variational Inference do not scale well to these conditions. Solutions may instead be found in Bayesian Last Layer approaches \cite{harrison2024variational}, where only the last layer of the Neural Network is treated as Bayesian. 

Alternatively, research on Mechanistic Interpretability, which analyses a large model's internal structure to gain insight into its decision making process \cite{bereska2024mechanistic}, may also offer avenues for uncertainty quantification. While Mechanistic Interpretability is not an uncertainty quantification method, early research is looking at using methods from Mechanistic Interpretability to find patterns and structures in foundation models that encode uncertainty \cite{cosentino2024reasoning}. Research on using Mechanistic Interpretability for Uncertainty Quantification is still in its early stages, but may become valuable when biosignal modelling starts to rely more on foundation models.

\subsection{Detecting Distribution Shift with Uncertainty}
Possibly the biggest risk in deploying Machine Learning systems in clinical settings is distribution shift. When a model is trained and evaluated on data from one setting, but is ultimately applied in a different setting the quality of the predictions will degrade. This may come from differences in the operators, recording hardware, patient populations or even bugs in the data handling. Such changes degrade the quality of predictions in ways that are typically not predictable.

Standard Machine Learning methods that only consider aleatoric uncertainty will be overconfident in these settings, while with epistemic uncertainty it may be possible to detect when the data distribution at deployment becomes different from the training distribution. 

Some effort on detecting distribution shift in biosignals exists, but no effective methods have been established \cite{xia2023benchmarking}. Additionally, current work is limited to synthetic shifts that might not represent shifts that would occur in reality. Datasets that combine recordings across clinics, contexts (inpatient vs outpatient), patient populations and time can show which distribution shifts occur and allow Machine Learning models to be trained in one context and thoroughly evaluated in another. Good epistemic uncertainty estimation should be able to estimate the lower quality of predictions under distribution shift, and research to establish this is duly needed.

\section{Conclusion}\label{sec:conclusion}

This review finds that Uncertainty Quantification methods for Neural Networks have been gaining increasing attention in the biosignal domain for the last five years, but that there are some hurdles to overcome. 

By providing clarification about how uncertainty measures relate to aleatoric and epistemic uncertainty, and by providing an end-to-end guideline on how to add uncertainty quantification to a biosignal classifying Neural Network we make uncertainty quantification more accessible to researchers working with EEG, ECG, EMG and EOG.

Many areas still remain to be explored. Uncertainty Quantification methods should be further studied in situ, where clinicians may perform specific actions based on predicted uncertainty. To this end, studies that investigate the performance of a (clinical) environment containing an uncertainty-estimating model are needed.

\section*{Acknowledgment}
The authors received no financial support for the research, authorship, and/or publication of this article.

\printbibliography

\end{document}